\documentclass[aps,preprint,prd,showpacs,superscriptaddress,amssymb,amsmath,preprintnumbers,floatfix]{revtex4-1}
\usepackage{graphicx}

\newcommand{\del}{\partial}
\newcommand{\td}{{\rm{d}}}
\newcommand{\e}{{\rm{e}}}
\newcommand{\img}{{\rm{i}}}

\begin{document}
\preprint{OU-HET-924, KEK-CP-356}
\title{
  Lattice calculation of coordinate-space
  vector and axial-vector current correlators in QCD
} 

\newcommand{\KEK}{
Theory Center,
Institute of Particle and Nuclear Studies,
High Energy Accelerator Research Organization (KEK),
Tsukuba 305-0801, Japan
}
\newcommand{\SOKENDAI}{
Department of Particle and Nuclear Physics,
SOKENAI (The Graduate University for Advanced Studies),
Tsukuba 305-0801, Japan
}
\newcommand{\Columbia}{
Physics Department,
Columbia University,
New York 10027, USA
}
\newcommand{\Edinburgh}{
School of Physics and Astronomy,
The University of Edinburgh,
Edinburgh EH9 3JZ, United Kingdom
}
\newcommand{\Osaka}{
Department of Physics,
Osaka University,
Toyonaka 560-0043, Japan
}

\author{M.~Tomii}
\email{mt3164_at_columbia.edu}
\affiliation{\Columbia}

\author{G.~Cossu}
\affiliation{\Edinburgh}

\author{B.~Fahy}
\affiliation{\KEK}

\author{H.~Fukaya}
\affiliation{\Osaka}

\author{S.~Hashimoto}
\affiliation{\KEK}
\affiliation{\SOKENDAI}

\author{T.~Kaneko}
\affiliation{\KEK}
\affiliation{\SOKENDAI}

\author{J.~Noaki}
\affiliation{\KEK}

\collaboration{JLQCD Collaboration}



\begin{abstract}
  We study the vector and axial-vector current correlators in
  perturbative and non-perturbative regimes of QCD.
  The correlators in Euclidean coordinate space
  are calculated on the lattice using the M\"obius
  domain-wall fermion formulation at three lattice spacings covering
  0.044--0.080~fm.
  The dynamical quark effects of $2+1$ light flavors are included.
  The sum $V+A$ and the difference $V-A$ of the vector ($V$)
  and axial-vector ($A$) current
  correlators calculated on the lattice after extrapolating to the
  physical point agree with those converted from the ALEPH experimental
  data of hadronic $\tau$ decays.
  The level of the agreement in the $V+A$ channel is about $1.3\sigma$
  or smaller in the region of $|x|\ge0.4$~fm, while that in the $V-A$ channel
  is about $1.8\sigma$ at $|x|=0.74$~fm and smaller at other distances.
  We also extract the chiral condensate from the short-distance correlators
  on the lattice using the PCAC relation. Its result extrapolated to the
  chiral and continuum limit is compatible with other estimates at low energies.
\end{abstract}

\maketitle

\section{Introduction}
\label{sec:intro}

The two-point current correlator is one of the most fundamental quantities
in the study of Quantum Chromodynamics (QCD).
It is defined as a vacuum expectation value of a product of quark
currents, and reflects the QCD dynamics.
It shows different features depending on the distance between the
currents. 
At short distances ($< 0.1$~fm), it behaves perturbatively,
{\it i.e.} the perturbative expansion about small coupling constant
works reasonably well.
Several properties including its scaling are
understood perturbatively.
In this region, the effect of spontaneous chiral symmetry breaking is
small and two correlators connected by the chiral transformation
become almost degenerate.
At long distances ($>1$~fm), on the other hand, current correlators
are saturated by the ground state and are characterized by its mass and
decay constant.
Degeneracy between the chiral partners is clearly lost.

In the distance region between the two regimes, $\sim$ 0.1--1~fm, neither the
perturbative nor the hadronic description is fully applicable.
Terms of higher powers in the QCD
coupling constant $\alpha_s(Q)$ become more significant, 
or the expansion even ceases to converge.
This is related to the emergence of power corrections through 
$e^{1/\beta_0\alpha_s(Q)}\sim(\Lambda_\mathrm{QCD}/Q)$
due to the running of the coupling as a function of the scale $Q$, an
inverse of the distance scale.
The QCD scale $\Lambda_\mathrm{QCD}$ characterizes the distance
scale where power corrections of the form
$(\Lambda_\mathrm{QCD}/Q)^n$ become important.
In the hadronic picture, this region is identified by many resonances
and multi-body scattering states.
The individual states involved are complicated, but the common belief
is that the sum over a number of hadronic states can be interpreted as
interacting quarks and gluons, {\it i.e.} quark-hadron duality.
There are many sources of evidence that this duality works, such as the
perturbative description of the experimentally measured $R$ ratio of
the $e^+e^-$ cross section, but theoretical understanding based on QCD
is as yet unsatisfactory.

A lattice QCD calculation is, in principle, applicable to any distance
scales in Euclidean space. 
So far, it has been successfully used to calculate hadron
correlators at long distances to extract hadron masses and matrix
elements, and precise agreement with experimental data for
many physical quantities is reported.
In such a calculation, the data at the short and middle distances are
ignored to avoid ``contamination'' from excited states, although they
may contain interesting information about the intermediate regime
where the quark and hadron pictures overlap.
In this work, we explore this regime using lattice data.

The vector and axial-vector current correlators can be related to 
hadronic $\tau$ decays and $e^+e^-$ hadronic cross section through the
optical theorem, which involves a weighted integral over the square of
the momentum transfer.
This connection between correlators and experimental data allows
us to compare the lattice calculation with experiment in the region
where excited states contribute significantly.

The vector and axial-vector current correlators in Euclidean space were
reconstructed using the early experimental data \cite{Shuryak:1993kg,Schafer:2000rv}.
They provide correlators in coordinate space with space-like
separation $x$ between the currents.
This is equivalent to space-like correlators in momentum space
after an appropriate Fourier transform, but allows more direct
comparison with the lattice calculation.
This sort of comparison was attempted previously using quenched simulations 
\cite{Chu:1993cn,Hands:1994cj,DeGrand:2001tm}.
We revisit this problem because there has been considerable progress in lattice
calculations and updated experimental data since.

In this work, we study correlators of the iso-triplet vector and  
axial-vector currents.
The most recent experimental data for these correlators are obtained 
through the hadronic $\tau$ decay experiment by the ALEPH
collaboration 
\cite{Davier:2013sfa}.
This experiment provides the spectral functions, which are functions of the
invariant mass $s$, with kinematical upper limits
set by the $\tau$ lepton mass $m_\tau^2$. 
Above this limit, the spectral function needs to be estimated using
perturbation theory, which is available
to the order of $\alpha_s^4$ \cite{Baikov:2008jh,Baikov:2009uw}
and reliable at sufficiently large invariant masses.
At lower invariant masses, the observed spectral
functions show significant deviation from the perturbative prediction 
due to a violation of quark-hadron duality
\cite{Shifman:1994yf,Blok:1997hs,Shifman:2000jv,Bigi:2001ys}.
The duality violation may be modeled using the Regge theory with
the large-$N_c$ assumption
\cite{Shifman:2000jv,Bigi:2001ys,Cata:2008ru,Boito:2012cr}.

Our lattice calculation is performed on 2+1-flavor QCD gauge ensembles.
We employ the M\"obius domain-wall fermion formulation
\cite{Brower:2004xi,Brower:2012vk} for both sea and valence quarks. 
Since discretization effects may become more significant at
distance scales below 1~fm,
we take the continuum limit using
ensembles with lattice spacing $a\simeq$ 0.080, 0.055 and 0.044~fm.
As a result, the correlators at distances larger than $\simeq$
0.4~fm are obtained with errors well under control.
The same set of gauge ensembles has been used for studying
heavy-light decay constants \cite{Fahy:2015xka},
$D$ meson semileptonic form factors \cite{Kaneko:2017sct}, 
the determination of the charm quark mass \cite{Nakayama:2016atf}, a
calculation of the chiral condensate \cite{Cossu:2016eqs} and the
$\eta'$ mass \cite{Fukaya:2015ara}.

We use local vector and axial-vector currents constructed with
M\"obius domain-wall fermions. 
Since these currents are not conserving, a finite renormalization is
needed. 
In our previous work \cite{Tomii:2016xiv}, we determined the
renormalization factor using correlators in the perturbative
regime based on the X-space renormalization procedure
\cite{Martinelli:1997zc,Gimenez:2004me,Cichy:2012is}.
That is, we determine the renormalization factor such that the lattice
correlators at short distances reproduce the  continuum perturbative
calculation available up to the order of $\alpha_s^4$ \cite{Chetyrkin:2010dx}
for massless quarks.
Since the chiral symmetry on the lattice is precisely maintained,
the renormalization factors of the vector and axial-vector
currents are identical.
The present work is a natural extension of the previous one, as the
deviation from the perturbative regime is the main concern.

Besides the comparison with experiment, 
we extract the chiral condensate from the vector and axial-vector
correlators.
This appears as the leading power correction to the correlators,
reflecting the spontaneous chiral symmetry breaking in QCD.
The extraction is based on the partially conserved axial current (PCAC) relation,
through which the derivative of the axial-vector correlator is directly related to
the chiral condensate \cite{Becchi:1980vz}.
With good control of the discretization effects, we are able to
determine the chiral condensate.
The result is consistent with
another determination from the spectral density of the Dirac operator
\cite{Cossu:2016eqs}.

This paper is organized as follows.
In Section~\ref{sec:second}, we discuss the vector and axial-vector
current correlators in the continuum theory, including the conversion
of the experimental data through a dispersion relation.
The relation between the chiral condensate and the axial-vector
correlator is also discussed.
In Section~\ref{sec:lat_calc}, we summarize our lattice setup and
describe the method to reduce the discretization effect in the
lattice correlators.
The comparison of the lattice data with experiment and the
extraction of the chiral condensate are shown in
Section~\ref{sec:result}. 
Section~\ref{sec:summary} concludes the paper with some discussions.

\section{Current correlators}
\label{sec:second}

In this work, we study two-point correlation functions of the
iso-triplet vector and axial-vector currents in Euclidean coordinate space,
\begin{equation}
  \Pi_{V,\mu\nu} (x) = \langle V_\mu(x)V_\nu(0)^\dag\rangle,
  \hspace{10mm}
  \Pi_{A,\mu\nu} (x) = \langle A_\mu(x)A_\nu(0)^\dag\rangle,
  \label{eq:def_correl}
\end{equation}
where the currents are defined by
\begin{equation}
  V_\mu(x) = \bar u(x)\gamma_\mu d(x),
  \hspace{10mm}
  A_\mu(x) = \bar u(x)\gamma_\mu\gamma_5d(x),
  \label{eq:def_currents}
\end{equation}
with up and down quark fields $u(x)$ and $d(x)$.
We also analyze the sum of the Lorentz diagonal components,
\begin{equation}
  \Pi_{V/A}(x) = \sum_\mu\Pi_{V/A,\mu\mu}(x).
  \label{eq:def_sum_va}
\end{equation}
In this work,
we take the masses of up and down quarks to be degenerate.

In the momentum space, the corresponding vacuum polarization tensors
$\widetilde\Pi_{V/A,\mu\nu}(Q)$ are given by
\begin{align}
  \widetilde\Pi_{V/A,\mu\nu}(Q)
  & = \int\td^4x\ \e^{-\img Qx}\ \Pi_{V/A,\mu\nu}(x)
    \notag\\
  & = (Q^2\delta_{\mu\nu}-Q_\mu Q_\nu)\widetilde\Pi_{V/A}^{(1)}(q^2)
    - Q_\mu Q_\nu\widetilde\Pi_{V/A}^{(0)}(q^2),
    \label{eq:VAcorr_mom}
\end{align}
where $\widetilde\Pi_{V/A}^{(J)}(q^2)$ is the vacuum polarization
function in the spin $J$ channel written as a function of the momentum
squared in Minkowski space, $q^2 = -Q^2$.

\subsection{Correlators from experiment}
\label{subsec:corr}

The vector and axial-vector correlators $\Pi_{V/A}(x)$ in
coordinate space are related to the experimental observables through
a dispersion relation.
In the momentum space, it is given by the well-known analyticity formula 
\begin{equation}
  \widetilde\Pi_{V/A}^{(J)}(q^2)
  = \frac{1}{\pi}\int_0^\infty\td s
  \frac{ {\rm Im}\ \widetilde\Pi_{V/A}^{(J)}(s) }{s-q^2}
  - \mbox{subtraction}.
  \label{eq:disp_rel0}
\end{equation}
Inserting this into (\ref{eq:VAcorr_mom}) and Fourier transforming
back to coordinate space, the correlators are found to be
\cite{Shuryak:1993kg,Schafer:2000rv}
\begin{equation}
  \Pi_{V/A}(x) = \frac{1}{8\pi^4}\int_0^\infty\td s\ s^{3/2}
  \left(3\rho_{V/A}^{(1)}(s) - \rho_{V/A}^{(0)}(s)\right)
  \frac{K_1(\sqrt{s}|x|)}{|x|},
  \label{eq:disp_rel_coord}
\end{equation}
where $K_1$ is the modified Bessel function and
$\rho_{V/A}^{(J)}(s) = 2\pi{\rm\ Im}~\widetilde\Pi_{V/A}^{(J)}(s)$ is 
the {\it so-called} spectral function.
The second term on the RHS of (\ref{eq:disp_rel0}), an unphysical
contact term, is proportional to $\delta(|x|)$ in coordinate space
and is therefore omitted from (\ref{eq:disp_rel_coord}) and in the
following discussions.

The spectral function represents the hadronic spectrum associated with
the corresponding current and spin $J$.
The spin-1 part $\rho_{V/A}^{(1)}(s)$ is measured by hadronic
$\tau$ decay experiments
\cite{Barate:1997hv,Barate:1998uf,Schael:2005am,Davier:2013sfa}.
The spin-0 part of the vector channel vanishes in the isospin
limit, while that of the axial-vector channel is dominated by the pion
pole,
$\rho_A^{(0)}(s) \propto f_\pi^2\delta(s-m_\pi^2)$ with 
$m_\pi$ and $f_\pi$ the pion mass and decay constant,
respectively.

Sch\"afer and Shuryak \cite{Schafer:2000rv} converted
$\rho_{V/A}^{(1)}(s)$ measured by 
ALEPH \cite{Barate:1997hv,Barate:1998uf} to the correlators
(\ref{eq:disp_rel_coord}), while the contribution of the spin-0
part $\rho_A^{(0)}(s)$ of the axial-vector channel was approximated by
using the mass and decay constant of the pion as explained above.
Their result was used to test consistency with a quenched
lattice simulation \cite{DeGrand:2001tm}.

In this work, we use the latest ALEPH data for $\rho_{V/A}^{(1)}(s)$
from $\tau$ decays \cite{Davier:2013sfa} to calculate
\begin{equation}
  \Pi_{V/A}^{(1)}(x) =
  \frac{3}{8\pi^4}\int_0^\infty\td s\ s^{3/2} 
  \rho_{V/A}^{(1)}(s) \frac{K_1(\sqrt{s}|x|)}{|x|},
\label{eq:disp_rel_spin1}
\end{equation}
which does not contain the contribution of the spin-0 part.
In
Section~\ref{subsec:Lat_vs_ALEPH_V+A}--\ref{subsec:Lat_vs_ALEPH_V-A},
we show the result of the
lattice calculation for $\Pi_{V/A}^{(1)}(x)$ extrapolated to the
physical point and discuss the consistency with experiment.

Since the spectral functions obtained from hadronic $\tau$ decays are
measured in a limited region of the invariant mass below the
$\tau$ lepton mass, $s<m_\tau^2$,
we need to complement this using some theory or model for the region
$s>m_\tau^2$ in order to estimate the integral in (\ref{eq:disp_rel_spin1}). 
The spectral functions are calculated through order $\alpha_s^4$ in
perturbation theory \cite{Baikov:2008jh,Baikov:2009uw}, which allows us to
precisely estimate the spectral functions in the perturbative regime.
Another possibility is to use the Operator Product Expansion (OPE)
technique \cite{Shifman:1978bx}, but it is known
that the OPE of the spectral functions in the Minkowski domain
disagrees with that in full QCD beyond the uncertainty due to the
truncation of the perturbative expansion and the operator expansion
\cite{Shifman:1994yf,Blok:1997hs,Shifman:2000jv,Bigi:2001ys}.
(Such disagreement is usually referred to as the quark-hadron duality
violation.)
Due to this, one needs to rely on models to estimate the spectral 
functions in the region beyond experimental reach.
Following a widely used model based on the Regge theory with a large-$N_c$
assumption
\cite{Shifman:2000jv,Bigi:2001ys,Cata:2008ru,Boito:2012cr},
we parametrize the spectral functions at large $s$ as
\begin{align}
  \rho_{V/A}^{(1)}(s) = \rho_{V/A}^{\rm pert}(s) +
  \e^{-\delta_{V/A}-\gamma_{V/A}s}
  \sin\left(\alpha_{V/A}+\beta_{V/A}s\right),
  \label{eq:spfunc_dv}
\end{align}
with the perturbative part $\rho^{\rm pert}_{V/A}(s)$
of the spectral functions \cite{Baikov:2008jh,Baikov:2009uw}
and unknown parameters 
$\delta_{V/A}$, $\gamma_{V/A}$, $\alpha_{V/A}$ and $\beta_{V/A}$.
The remnant of resonances appears as the oscillatory term, which  is
exponentially suppressed at higher energies.

We perform a global fit for the vector and axial-vector spectral
functions measured by ALEPH \cite{Davier:2013sfa} to determine the
unknown parameters
$\delta_{V/A}$, $\gamma_{V/A}$, $\alpha_{V/A}$ and $\beta_{V/A}$ 
taking account of the correlation between these two channels.
We choose the fit range 
$1.6{\rm\ GeV^2} \le s \le 2.7{\rm~GeV^2}$,
in which the fit function (\ref{eq:spfunc_dv}) is supposed to be
valid and the statistical uncertainty of the experimental data is
not too large.
As a result of the global fit, we obtain the parameters as
\begin{align}
  \delta_V = 0.32(27),\ \ \gamma_V = 0.72(9){\rm~GeV}^{-2},
  \ \ 
  \alpha_V = -2.4(9),\ \ \beta_V = 4.3(2){\rm~GeV}^{-2},
  \notag\\
  \delta_A = -1.5(5),\ \ \gamma_A = 1.7(2){\rm~GeV}^{-2},
  \ \ 
  \alpha_A = 2.2(4.8),\ \ \beta_A = 3.6(1.2){\rm~GeV}^{-2}.
  \label{eq:fit_DVmodel}
\end{align}
There are 24 degrees of freedom and the value of $\chi^2$ per degree of
freedom is 1.3.

\begin{figure}[tbp]
  \begin{center}
    \includegraphics[width=120mm]{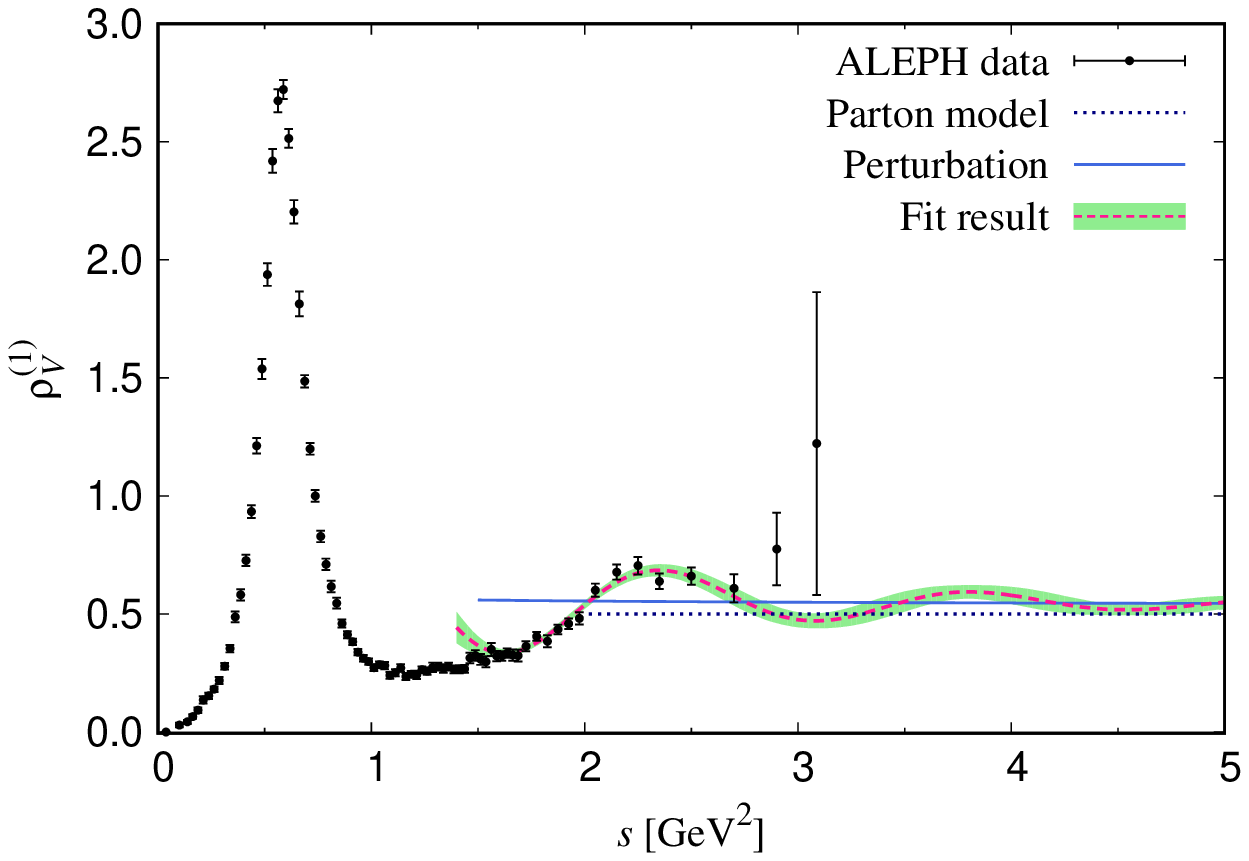}
  \end{center}
  \vspace{6mm}
  \begin{center}
    \includegraphics[width=120mm]{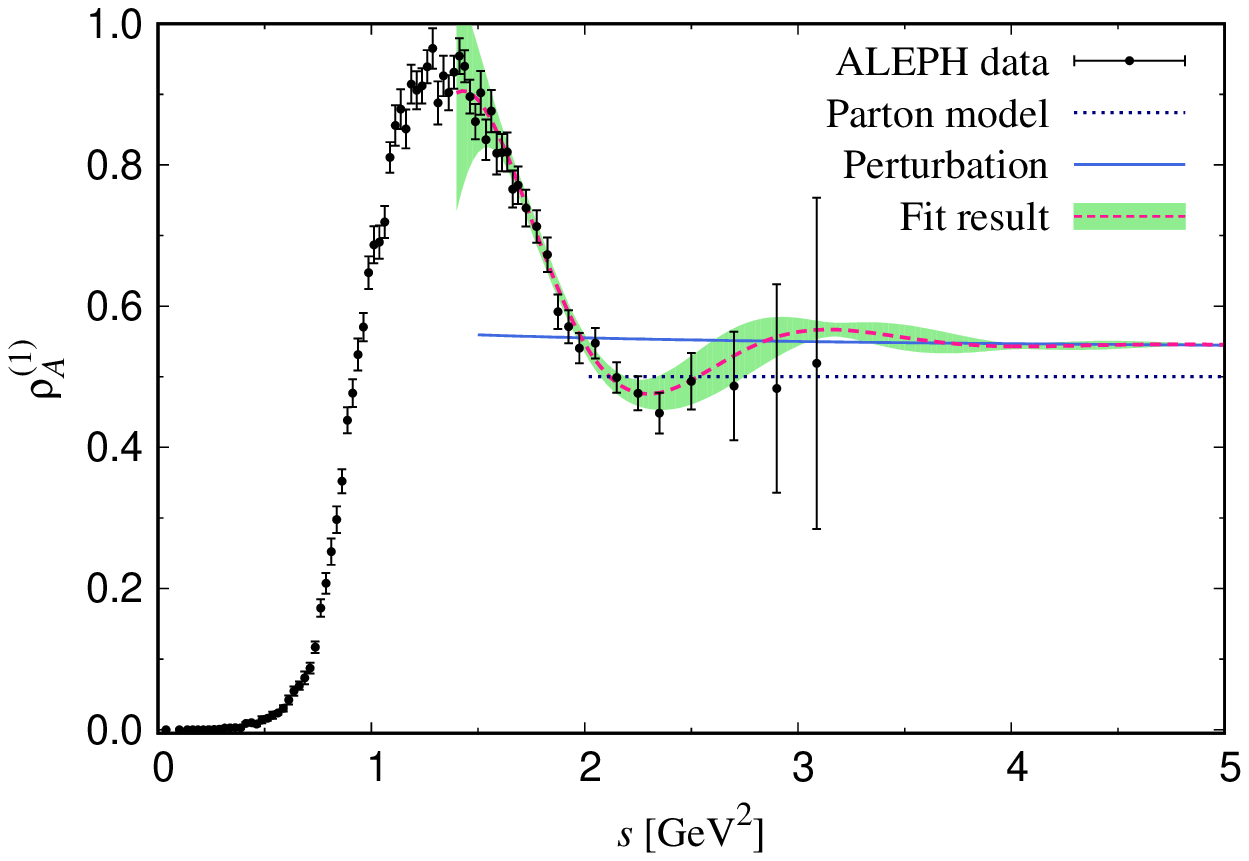}
  \end{center}
  \caption{
    Spectral functions of the vector (upper) and axial-vector (lower)
    channels measured by the ALEPH collaboration (circles)
    \cite{Davier:2013sfa} as functions of $s$.
    The prediction of the parton model (dotted line), perturbation
    theory (solid line) and the fit result (dashed curve and band)
    using the fit function (\ref{eq:spfunc_dv}) are also shown.
  }
  \label{fig:spfunc}
\end{figure}

Figure~\ref{fig:spfunc} shows the spectral functions in the vector
(upper panel) and axial-vector (lower panel) channels measured by
ALEPH \cite{Davier:2013sfa}.
The dotted and solid lines stand for the prediction of the parton
model (corresponding to the leading order perturbation theory) and
the perturbation theory at $O(\alpha_s^4)$, respectively.
The fit result is represented by the dashed curve and the band.
For the vector channel, the effect of the duality violation is visible
as a bump around $s\simeq$ 2.5~GeV$^2$.
In order to converge towards the perturbative prediction at high
energies, the oscillatory and decaying function of the form
$\e^{-\delta_{V/A}-\gamma_{V/A}s} \sin(\alpha_{V/A}+\beta_{V/A}s)$
is necessary.

\begin{figure}[tbp]
  \begin{center}
    \includegraphics[width=120mm]{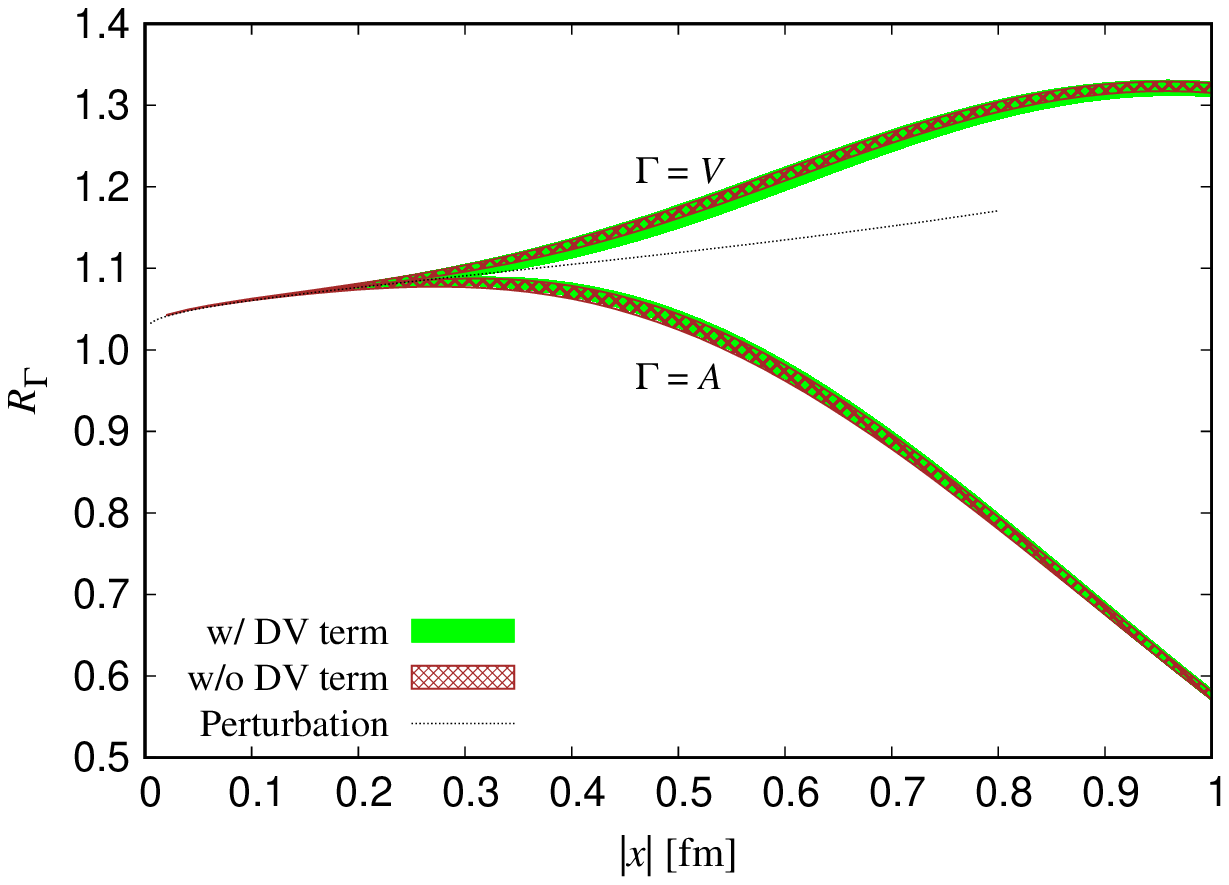}
    \caption{
      Vector and axial-vector correlators reproduced using the
      dispersion relation (\ref{eq:disp_rel_spin1}).
      The spectral functions measured by ALEPH are used for 
      $s \le 2.7\rm~GeV^2$, while those in $s > 2.7\rm~GeV^2$ are
      calculated perturbatively with (solid band) 
      and without (hatched band) the duality-violating term.
      The prediction of the massless perturbation theory is also shown
      (dotted line). 
    }
    \label{fig:v1_a1_cont}
  \end{center}
\end{figure}

The correlators reconstructed using
(\ref{eq:disp_rel_spin1}) are shown in Figure~\ref{fig:v1_a1_cont}.
They are normalized by the tree-level value
$R_{V/A}(x) = \Pi_{V/A}^{(1)}(x)/\Pi_0(x)$ 
with $\Pi_0(x)$ the correlator in the massless free
theory, which is the same for the vector and axial-vector channels.
We divide the integral (\ref{eq:disp_rel_spin1})
into two regions at $s_0=2.7\rm~GeV^2$.
Below $s_0$, we directly input the spectral functions from experiment.
Above $s_0$, the spectral functions from massless
perturbation theory with (solid band) and without (hatched band) the
duality-violating term are used.
There are two remaining experimental data points for each channel above $s_0$
that has been discarded in this analysis due to the large statistical errors.
At short distances ($<0.2$~fm), these correlators agree with the prediction of
massless perturbation theory (dotted line) \cite{Chetyrkin:2010dx}.

One can see that the impact of the duality-violating term is not very significant.
This is reasonable because the effect of the duality violation is
smeared out by the dispersion integral.
The vacuum polarization function in the space-like region is
insensitive to the individual poles in the Minkowski domain.
We show consistency between these results with the lattice calculation
in Section~\ref{subsec:Lat_vs_ALEPH_V+A}--\ref{subsec:Lat_vs_ALEPH_V-A}.

\begin{figure}[pbp]
  \begin{center}
    \includegraphics[width=120mm]{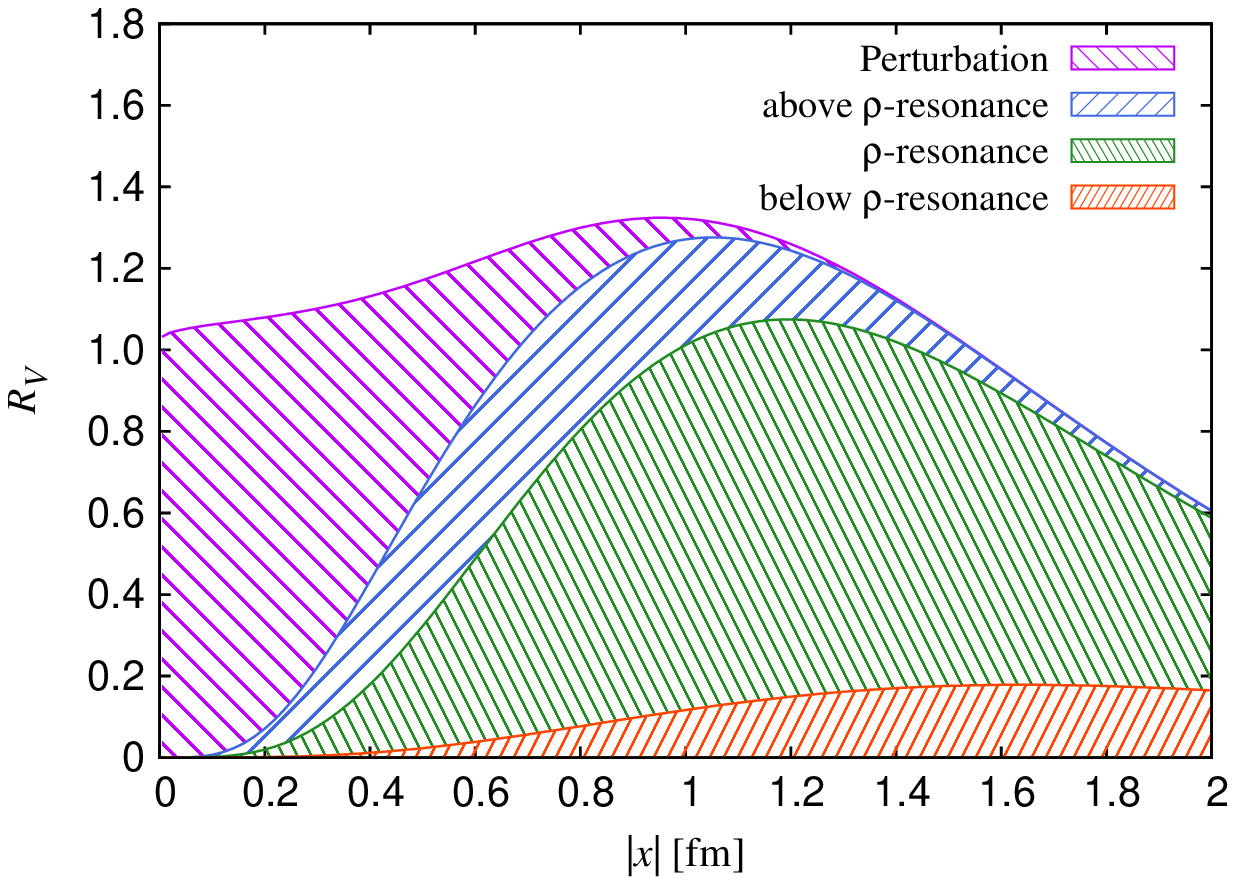}
  \end{center}
  \caption{
    Decomposition of the vector correlators into contributions
    from the spectral function in different regions of $s$.
  }
  \label{fig:dominance_v1}
  \vspace{6mm}
  \begin{center}
    \includegraphics[width=120mm]{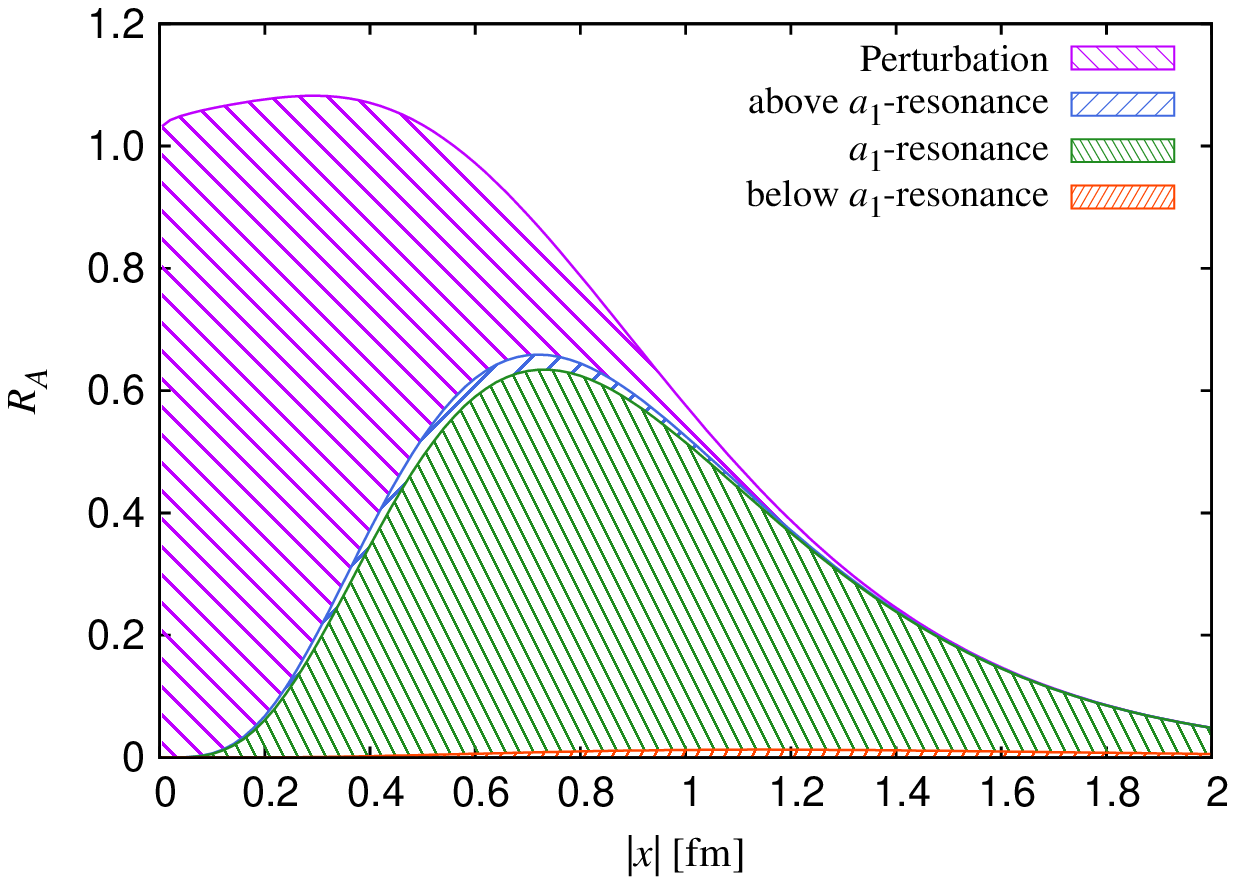}
  \end{center}
  \caption{
    Same as Figure~\ref{fig:dominance_v1} but for the axial-vector channel.
  }
  \label{fig:dominance_a1}
\end{figure}

It is convenient for later discussions
to investigate the size of non-perturbative contributions to the
correlators at each distance $|x|$.
Figure~\ref{fig:dominance_v1} shows the decomposition of $R_V(x)$ into 
contributions from the spectral function in different regions of $s$.
The area indicated by ``Perturbation'' represents the contribution from
the spectral function for $s>2.7\rm~GeV^2$, 
which is calculated perturbatively.
For the more non-perturbative regime, we split the region $s\le2.7\rm~GeV^2$
into three: the $\rho$ meson resonance
$(0.776 -0.150)^2{\rm~GeV}^2 < s < (0.776 + 0.150)^2~{\rm GeV}^2$,
plus the regions above and below it.
We also show the corresponding plot for the axial-vector channel in
Figure~\ref{fig:dominance_a1}.
The region of $s$ indicated by ``Perturbation'' is the same as for the
vector channel, {\it i.e.} $s > 2.7~\rm GeV^2$, 
while the resonance of the $a_1$ meson is chosen as
$(1.23 - 0.40)^2~{\rm GeV}^2 < s < (1.23 + 0.40)^2~{\rm GeV}^2$.
Both plots indicate that the non-perturbative effect is quite
significant in the distance region around $|x| \simeq 0.5$~fm, 
but the correlators are not saturated by the ground state.
This is the region that we are interested in, {\it i.e.} neither the
perturbative expansion nor low-energy effective theories are fully applicable.
In Section~\ref{subsec:Lat_vs_ALEPH_V+A}--\ref{subsec:Lat_vs_ALEPH_V-A},
we demonstrate that the lattice calculation succeeds in reproducing the
experimental results at $|x|\simeq0.5$~fm.

\subsection{Chiral condensate through PCAC relation}
\label{subsec:PCAC_efcc}

While the spin-1 part of the vector and axial-vector correlators is
related to the hadronic $\tau$ decays as discussed in the previous
subsection, their spin-0 part is sensitive to another feature of
QCD. 
In the isospin limit, the spin-0 part of the vector channel
vanishes, so only the axial-vector channel is non-trivial.

According to the PCAC relation, projection of the axial-vector
correlator to the spin-0 part is achieved by taking its
divergence. 
Using the PCAC relation, we can relate the axial-vector correlator to 
the chiral condensate $\langle \bar qq\rangle$ as follows.
The spin-0 part of the axial-vector correlator is given by
\cite{Becchi:1980vz,Jamin:1992se}
\begin{equation}
  \widetilde\Pi_A^{(0)}(q^2)
  = \frac{4m_q}{q^4}\langle\bar qq\rangle
  + \frac{4m_q^2}{q^4}\widetilde\Pi_P(q^2),
\end{equation}
where $m_q$ is the degenerate mass of up and down quarks and
\begin{equation}
  \widetilde\Pi_P(q^2) = \int\td^4x\ \e^{-\img Qx}
  \langle\bar{u}\img\gamma_5d(x)
  \cdot
  \bar{d}\img\gamma_5u(0)\rangle.
\end{equation}
Therefore, the Fourier transform of (\ref{eq:VAcorr_mom}) leads to 
\begin{equation}
  -\frac{\pi^2}{2m_q}x^2\sum_{\mu,\nu}x_\nu\del_\mu\Pi_{A,\mu\nu}(x)
  = -\langle\bar qq\rangle
  + O(m_q/x^2).
  \label{eq:AWI_cont}
\end{equation}
Here, the renormalization scheme and scale dependence of $m_q$ account
for those of $\langle\bar qq\rangle$.
Taking the chiral limit of (\ref{eq:AWI_cont}), we can extract
the chiral condensate 
$\Sigma = -\lim_{m_q\rightarrow0}\langle\bar qq\rangle$.
While the dependence on mass and $|x|$ is $O(m_q/x^2)$ at short distances,
this quantity decreases exponentially at long distances,
$\sim \e^{-m_\pi|x|}$. This is discussed in Section~\ref{subsec:efcc},
where we calculate the chiral condensate based on this relation.

\section{Lattice calculation}
\label{sec:lat_calc}

\tabcolsep = 6pt
\begin{table}[tb]
\caption{
Lattice ensembles used in this work.
}
\label{tab:ensembles}
\begin{center}
\begin{tabular}{ccccccccc}
\hline
\hline
$\beta$ & $a$ [fm] & $N_s^3\times N_t\times L_s$ & $am_s$
& $am_q$ & $am_{res}$ & $aM_\pi$ & $N_{\rm conf}$ & $N_{\rm src}$\\
\hline
4.17 & 0.0804 & $32^3\times64\times12$ & 0.0300 & 0.0070 & 0.00017(1) & 0.1263(4) & 200 & 4\\
& & & & 0.0120 & 0.00015(2) & 0.1618(3) & 200 & 2 \\
& & & & 0.0190 & 0.00015(3) & 0.2030(3) & 200 & 2 \\
& & $48^3\times96\times12$ & 0.0400 & 0.0035 & 0.00022(2) & 0.0921(1) & 200 & 2 \\
& & $32^3\times64\times12$ & & 0.0070 & 0.00023(4) & 0.1260(4) & 200 & 4 \\
& & & & 0.0120 & 0.00012(8) & 0.1627(3) & 200 & 2 \\
& & & & 0.0190 & 0.00015(3) & 0.2033(3) & 200 & 2 \\
\hline
4.35 & 0.0547 & $48^3\times96\times8$ & 0.0180 & 0.0042 & $\sim 10^{-5}$ & 0.0820(3) & 200 & 2 \\
& & & & 0.0080 & & 0.1127(3) & 200 & 1 \\
& & & & 0.0120 & & 0.1381(3) & 200 & 1 \\
& & & 0.0250 & 0.0042 & & 0.0831(4) & 200 & 2 \\
& & & & 0.0080 & & 0.1130(3) & 200 & 1 \\
& & & & 0.0120 & & 0.1387(3) & 200 & 1 \\
\hline
4.47 & 0.0439 & $64^3\times128\times8$ & 0.0150 & 0.0030 & & 0.0632(2) & 200 & 1 \\
\hline
\hline
\end{tabular}
\end{center}
\end{table}

In this work, we use the lattice ensembles generated with $2+1$-flavor
dynamical M\"obius domain-wall fermions 
\cite{Brower:2004xi,Brower:2012vk}.
The tree-level improved Symanzik action \cite{Luscher:1984xn} is used
for the gauge part and the fermions couple to the gauge links after
three steps of the stout smearing \cite{Morningstar:2003gk}.
The gauge ensembles used in this analysis are summarized in
Table~\ref{tab:ensembles}.

The lattice spacing $a$ ranges between 0.044 and 0.080~fm, with which
we take the continuum limit.
Their values are determined through the Wilson-flow scale
$t_0^{1/2}$ \cite{Luscher:2010iy} with an input 
$t_0^{1/2}$ = 0.1465(21)(13)~fm taken from 
\cite{Borsanyi:2012zs}.
Degenerate up and down quark masses $m_q$ cover a range of
pion masses between 230 and 500~MeV.
The same masses are used for both sea and valence quarks.
The strange quark is, on the other hand, only in the sea, and its mass
$m_s$ is taken close to the physical value.
The residual mass $m_{res}$ is $O(1)$~MeV on the coarsest lattice and
much smaller than that on finer lattices.
For each ensemble, $N_{\rm conf}=200$ configurations are sampled from
5,000 molecular dynamics time.
On each configuration, we calculate the correlators with one or more
($N_{\rm src}$) source points.
We use the IroIro$++$ simulation code \cite{Cossu:2013ola} for these calculations.

We calculate the current-current correlators 
$\Pi_{V/A,\mu\nu}(x)$ in (\ref{eq:def_correl})
with the local vector and axial-vector currents defined on the lattice
using M\"obius domain-wall fermions.
The autocorrelation of correlators exists and is estimated to affect $O(10)$
nearby measurements on the finest lattice, which is examined
by varying bin size of jackknife samples. Among 200 configurations
analyzed, roughly 20 measurements are statistically
independent. The autocorrelation time of the topological charge is
about 4 times larger, but we do not find any significant correlation
with correlators.

We average correlators that are related by $90^\circ$ rotations.
In this way lattice points of different orientations are
distinguished even when they have the same $x^2$.
Namely, the points equivalent to the coordinate (1,1,1,1) are
distinct from (2,0,0,0), since they receive different
discretization effects.

As discussed in the previous paper \cite{Tomii:2016xiv}, we 
apply some cuts and corrections to reduce the discretization effects.
First, we subtract the dominant discretization effect by subtracting
the correlators constructed from lattice quarks in the free field theory
and add back their continuum counterparts.
This procedure is further improved by using the mean-field
approximation instead of the free propagator \cite{Lepage:1992xa}.
In addition,
we discard the data points that are expected to have large remaining discretization effects.
The criterion for the cut is given by an angle $\theta$ between the
position vector $x$ and the direction (1,1,1,1).
Since the lattice data at large $\theta$ tend to have large
discretization effects \cite{Chu:1993cn,Cichy:2012is}, 
we neglect the lattice data with $\theta \ge 30^\circ$.
This particular value is chosen such that
the points with a same $x^2$ become consistent within the statistical
error. 
More details are described in \cite{Tomii:2016xiv}.

\section{Results}
\label{sec:result}

\subsection{Consistency of the lattice data with ALEPH in the $V+A$ channel}
\label{subsec:Lat_vs_ALEPH_V+A}

In this subsection, we discuss the consistency between the correlators
calculated on the lattice and those converted from the ALEPH data for
hadronic $\tau$ decays.
The conversion of the experimental data is discussed in Section~\ref{subsec:corr}.
Here, we analyze the sum and difference of the vector and axial-vector
correlators, {\it i.e.} the $V+A$ and $V-A$ channels, normalized by
the corresponding free correlator $\Pi_0(x)$ in the massless limit
\begin{equation}
  R_{V\pm A}(x)
  = \frac{\Pi_V^{(1)}(x) \pm \Pi_A^{(1)}(x)}{2\Pi_0(x)}.
  \label{eq:Rvpma}
\end{equation}

For the lattice calculation, the vector and axial-vector currents need
to be renormalized since the local currents we use in this work are
not conserving.
The renormalization is done multiplicatively, {\it i.e.}
$R_{V\pm A}(x)\rightarrow {Z_V^{\rm\overline{MS}}(a)}^2R_{V\pm A}(x)$,
with $Z_V^{\rm\overline{MS}}(a)$ the renormalization factor of
the vector and axial-vector currents determined in the previous
work \cite{Tomii:2016xiv}.

Since the quantities $R_{V\pm A}(x)$ do not contain the spin-0
contribution, 
we also need to subtract the spin-0 part from the lattice correlators 
to obtain the spin-1 piece
$\Pi_{V/A}^{(1)}(x) = \Pi_{V/A}(x) - \Pi_{V/A}^{(0)}(x)$.
For the vector channel, the spin-0 part is absent in the isospin
limit, $\Pi_V^{(0)}(x)=0$.
For the axial-vector channel, we approximate the spin-0 part
$\Pi_A^{(0)}(x)$ by the contribution from the ground state pion.
In the infinite volume, it is given by
\begin{align}
  \Pi_A^{(0),\infty}(x)
  \simeq \frac{z_0M_\pi^2}{2\pi^2} \frac{K_1(M_\pi|x|)}{|x|},
\end{align}
where $z_0$ and $M_\pi$ are extracted from the zero-momentum
correlator
$\int\td^3x\ \Pi_{A,44}(\vec x,t) \rightarrow z_0\e^{-M_\pi t}$
at large time separations.
We neglect the excited states of the pion because they are not
expected to give significant contributions.
In a finite box, finite volume effects due to the pion pole may
appear. 
To take account of this, we subtract the wrap-around effect of the
pion and analyze
\begin{equation}
  \Pi_A^{(1)}(x) = \Pi_A(x) - \sum_{x_0}\Pi_A^{(0),\infty}(x-x_0),
\end{equation}
where the sum over $x_0$ runs over
\begin{equation}
x_0 \in \{(0,0,0,0),(\pm L,0,0,0),(0,\pm L,0,0),(0,0,\pm L,0),(0,0,0,\pm T),(\pm L,\pm L,0,0),\ldots\}.
\end{equation}

\begin{figure}[tbp]
  \begin{center}
    \includegraphics[width=120mm]{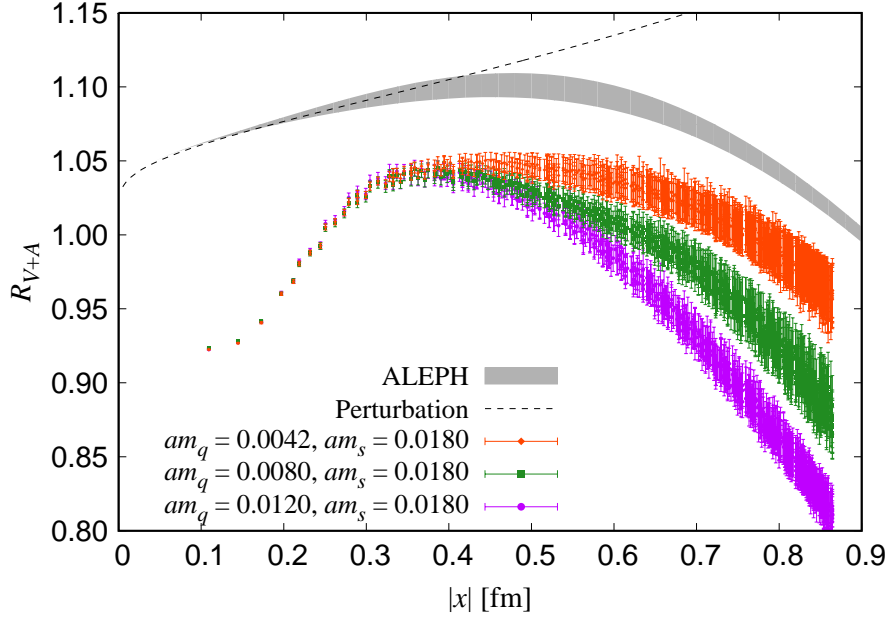}
    \caption{
      $R_{V+A}$ calculated on the ensembles with 
      $\beta$ = 4.35, $am_s$ = 0.0180
      and three input light quark masses:
      $am_q$ = 0.0042 (diamonds), 0.080 (squares) and 0.0120 (circles).
      The prediction of massless perturbation theory (dashed curve),
      and the results from experiment calculated with the
      duality-violating term in (\ref{eq:spfunc_dv}) (band) are also
      shown. 
    }
    \label{fig:vpa_ms0.0180}
  \end{center}
\end{figure}

\begin{figure}[tbp]
  \begin{center}
    \includegraphics[width=120mm]{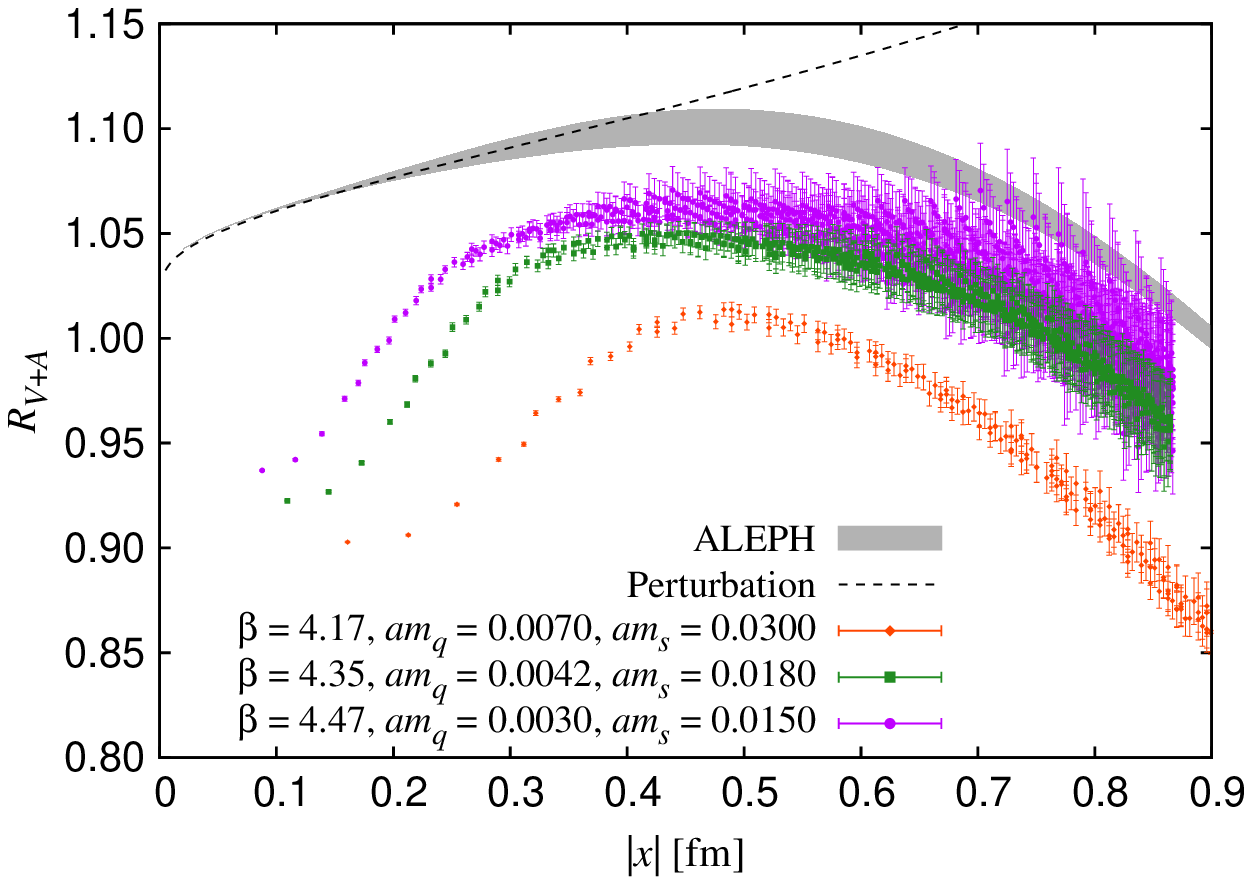}
    \caption{
      Same as Figure~\ref{fig:vpa_ms0.0180} but calculated on
      ensembles with different lattice cutoffs.  Pion masses
      are $M_\pi\simeq$ 300~MeV. 
    }
    \label{fig:vpa_3beta}
  \end{center}
\end{figure}

Figure~\ref{fig:vpa_ms0.0180} shows the results for $R_{V+A}(x)$ on
the ensembles with $\beta=4.35$, $am_s=0.0180$ and with different light
quark masses, $am_q$ = 0.042, 0.080 and 0.0120.
Here, we also show the prediction of massless perturbation theory
\cite{Chetyrkin:2010dx}
(dashed line) and the experimental result (band) from the dispersion
relation, where the spectral functions $\rho_{V/A}(s)$ 
in $s>2.7\rm~GeV^2$ include the duality-violating term in
(\ref{eq:spfunc_dv}).
In Figure~\ref{fig:vpa_ms0.0180}, one can see that the results at
smaller masses are closer to the experimental result.

Figure~\ref{fig:vpa_3beta} shows the results at pion masses
$M_\pi\simeq300$~MeV but with different $\beta$'s.
We find a significant dependence on the lattice spacing,
which can be described by the leading term, which is proportional to $a^2$ at middle
and long distances.
As we approach the continuum limit, the lattice data tend to approach
the experimental data.

We extrapolate these lattice results to the physical point, 
{\it i.e.} the continuum limit $a\to 0$ and physical pion mass $m_\pi\simeq140$~MeV.
To do so, we first divide the range of $|x|$ into $N$ bins,
\begin{align}
  B_i &= [x_i-\delta x/2,x_i+\delta x/2],\ \ x_{i+1} = x_i+\delta x,\ \ i
        = 1,2,\ldots, N,
\end{align}
where $x_i$ and $\delta x$ are the center of the $i$th bin
and the width of the bins, respectively.
For each bin, we average $R_{V+A}(x)$ over lattice points $x$ in $B_i$.
Since the average depends on the lattice spacing, input mass and
the representative distance $x_i$ of the correlator, the average is
denoted by $\overline R_{V+A}(a,M_\pi,x_i)$ with the explicit
dependence on these parameters.
Here, we neglect the dependence on the strange quark mass because
we do not find significant dependence in the lattice results. 
We then perform a global fit for all ensembles using the fit function 
\begin{equation}
  \overline R_{V+A}(a,M_\pi,x_i)
  = R_{V+ A}(0,m_\pi,x_i) + c_i(M_\pi^2-m_\pi^2) + d_i a^2,
  \label{eq:extrap}
\end{equation}
with three free parameters $R_{V+A}(0,m_\pi,x_i), \,c_i$ and $d_i$
for each $i$.
The first parameter $R_{V+A}(0,m_\pi,x_i)$ corresponds to the extrapolated value.
The other parameters $c_i$ and $d_i$ control the dependences on
the pion mass and the lattice spacing, respectively.

\begin{figure}[tbp]
  \begin{center}
    \includegraphics[width=120mm]{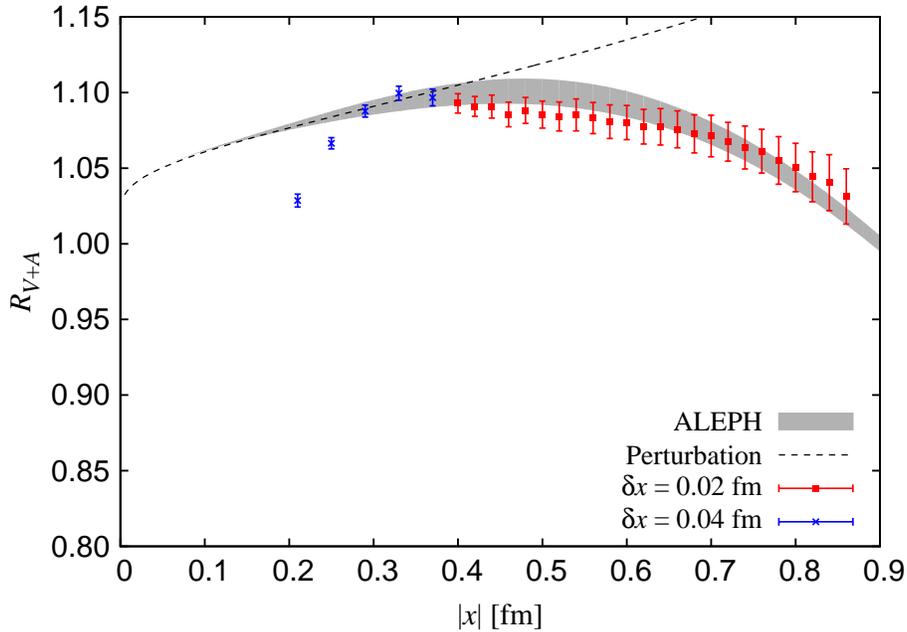}
    \caption{
      Lattice result for $R_{V+A}$ after the chiral and continuum
      extrapolations. 
      Data in each bin are extrapolated assuming (\ref{eq:extrap}).
      The bin size is larger in the short-distance region $|x|\lesssim$
      0.4~fm (blue crosses) than others (red squares) as there are
      fewer lattice points.
    }
    \label{fig:ALEPHvpa_ext}
  \end{center}
\end{figure}

\begin{figure}[tbp]
  \begin{center}
    \includegraphics[width=120mm]{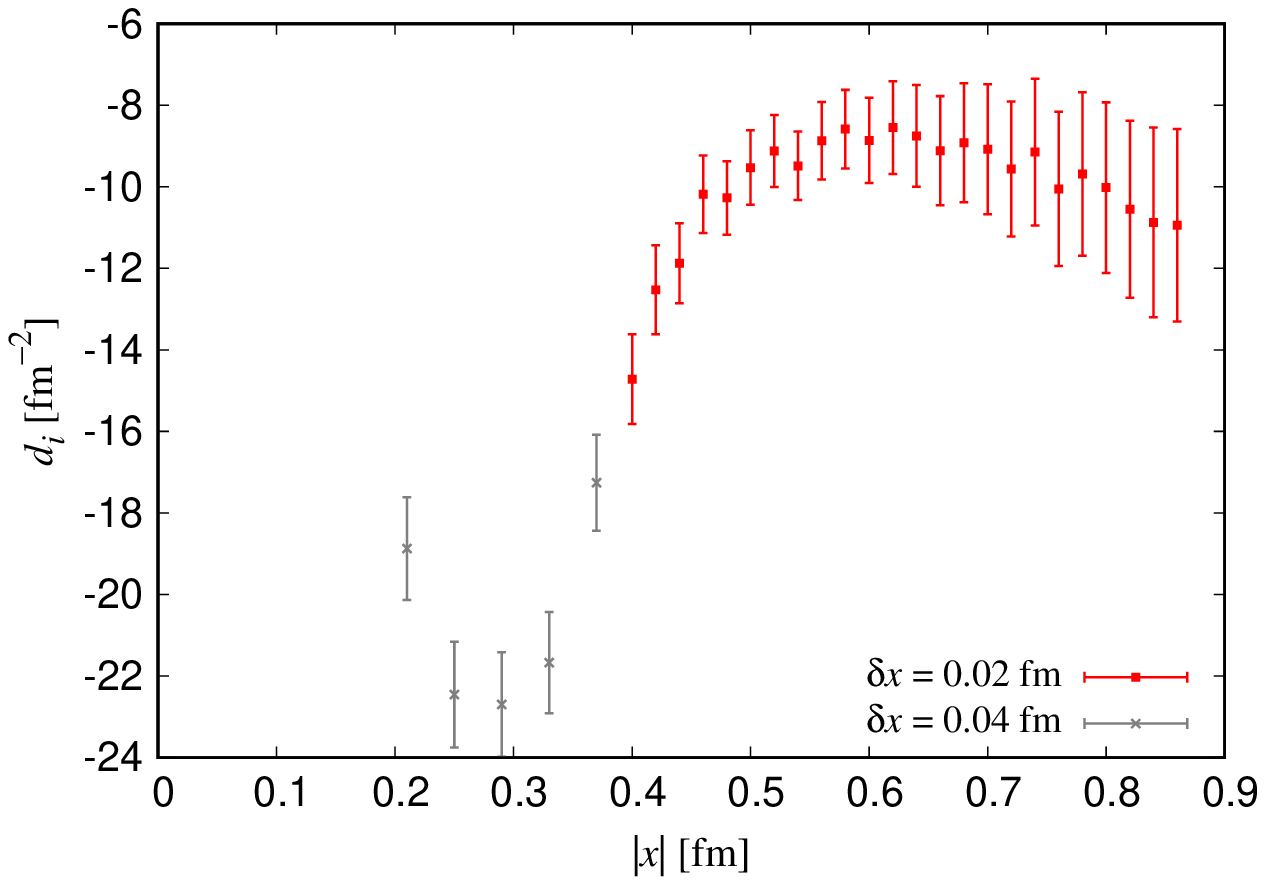}
  \end{center}
  \caption{
    Fit parameter $d_i$ obtained from the extrapolation of $R_{V+A}(x)$
    to the physical point.
    The result for each bin is plotted as a function of $|x| = x_i$.
  }
  \label{fig:ca_vpa}
  \vspace{6mm}
  \begin{center}
    \includegraphics[width=120mm]{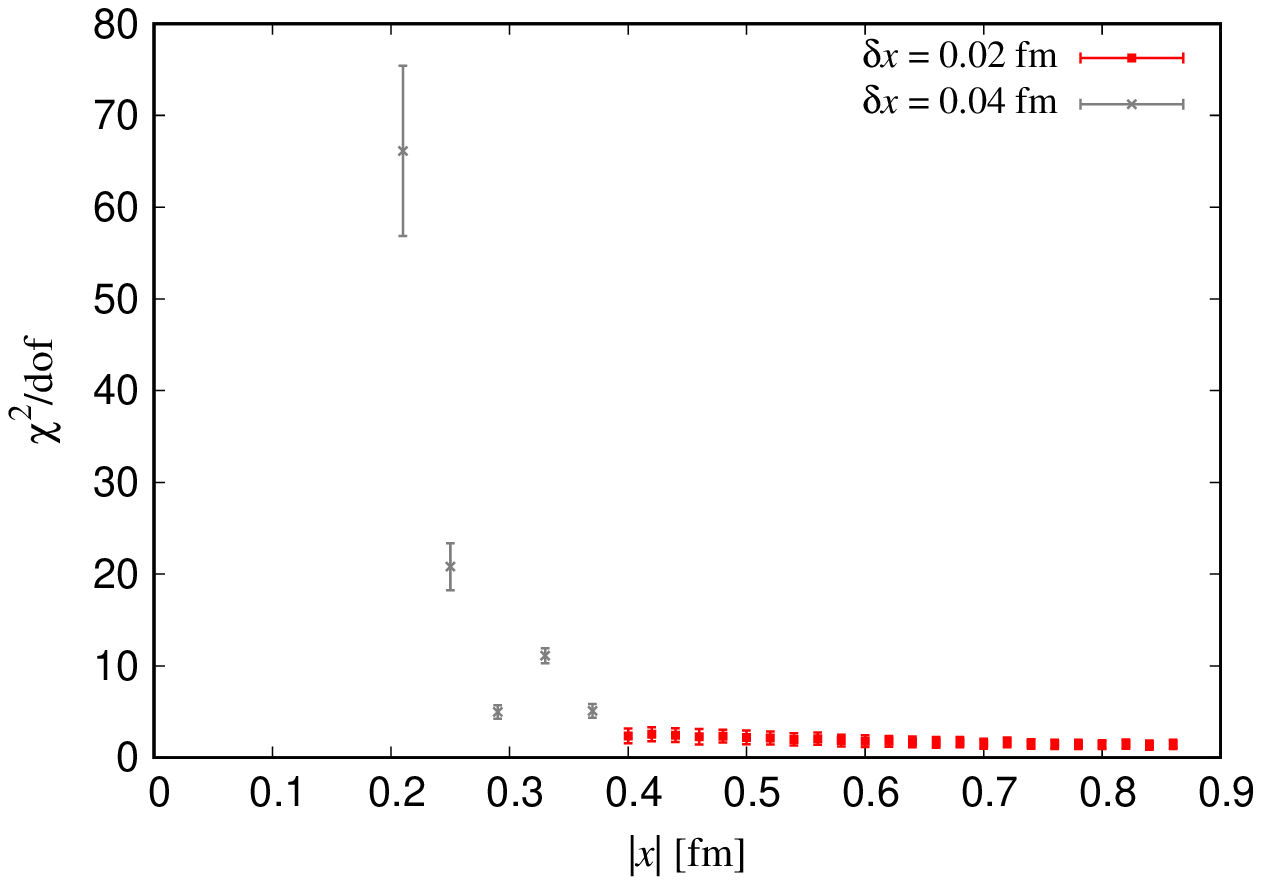}
  \end{center}
  \caption{
    $\chi^2$ divided by the degrees of freedom for the extrapolation
    of $R_{V+A}(x)$ to the physical point.
    The result for each bin is plotted as a function of $|x| = x_i$.
  }
  \label{fig:chi2_vpa}
\end{figure}

Figure~\ref{fig:ALEPHvpa_ext} shows the result of the extrapolation.
Here, we take $\delta x = 0.02$~fm for $x_i\ge0.4$~fm and 
$\delta x = 0.04$~fm 
for $x_i < 0.4$~fm.
While the agreement between the lattice result and
experiment is found for $|x| > 0.3$~fm in the plot, the fit function
\eqref{eq:extrap} may not be appropriate at short distances
due to remnant discretization effects that
could not be removed by the extrapolation linear in $a^2$.
In order to clarify the appropriate region of the fit function,
we show the fit result of the coefficient $d_i$ in Figure~\ref{fig:ca_vpa}
and $\chi^2/$dof in Figure~\ref{fig:chi2_vpa}.
In Figure~\ref{fig:ca_vpa}, the coefficient $d_i$ varies quite rapidly
at $|x|\simeq0.4$~fm, while it is mostly constant at longer distances.
As Figure~\ref{fig:chi2_vpa} shows, $\chi^2$/dof in $|x|<0.4$~fm is
much larger than that in the longer distance regime.
These results imply that the remnant discretization effects at $O(a^4)$
are not negligible in $|x| < 0.4$~fm.

\begin{figure}[tbp]
  \begin{center}
    \includegraphics[width=120mm]{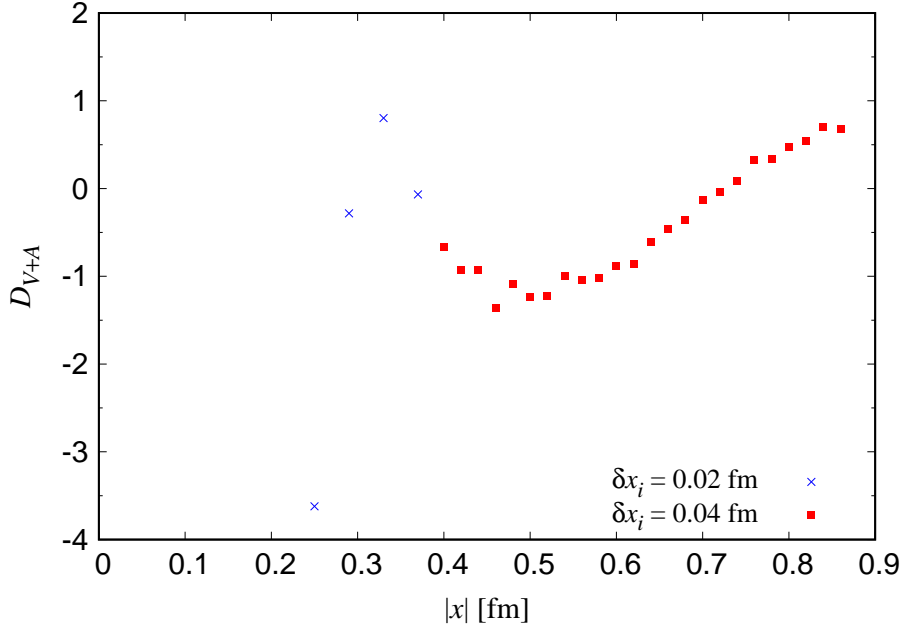}
    \caption{
	Significance of the difference $D_{V+A}(x)$ between the lattice calculation
	and experiment plotted as a function of $|x|$.
    }
    \label{fig:vpa_sigma}
  \end{center}
\end{figure}

To clarify the level of agreement between the lattice calculation
and experiment, we analyze the significance of the difference
\begin{equation}
D_{V\pm A}(x) = \frac{R_{V\pm A}^{\rm lat}(x)-R_{V\pm A}^{\rm exp}(x)}
{\sqrt{\delta {R_{V\pm A}^{\rm lat}(x)}^2+\delta {R_{V\pm A}^{\rm exp}(x)}^2}},
\end{equation}
where $R_{V\pm A}^{\rm lat (exp)}(x)$ is the central value of $R_{V\pm A}(x)$
calculated on the lattice (converted from experiment) in
Figure~\ref{fig:ALEPHvpa_ext} and $\delta R_{V\pm A}^{\rm lat (exp)}(x)$ is its
statistical error.
Figure~\ref{fig:vpa_sigma} shows $D_{V+A}(x)$ and indicates that
the lattice calculation agrees with experiment within $1.3\sigma$
in the region $|x|\ge0.4$~fm.

We may conclude that the lattice QCD calculation successfully
reproduces the spectral function observed in experiment at length
scale of 0.4~fm and larger.
The agreement around $|x|\sim0.5$~fm is important since the precise 
calculation of the correlators in this region is difficult with
perturbative approaches and with low-energy effective theories
as discussed in Section~\ref{subsec:corr}.

\subsection{Consistency of the lattice data with ALEPH in the $V-A$ channel}
\label{subsec:Lat_vs_ALEPH_V-A}

\begin{figure}[tbp]
  \begin{center}
    \includegraphics[width=120mm]{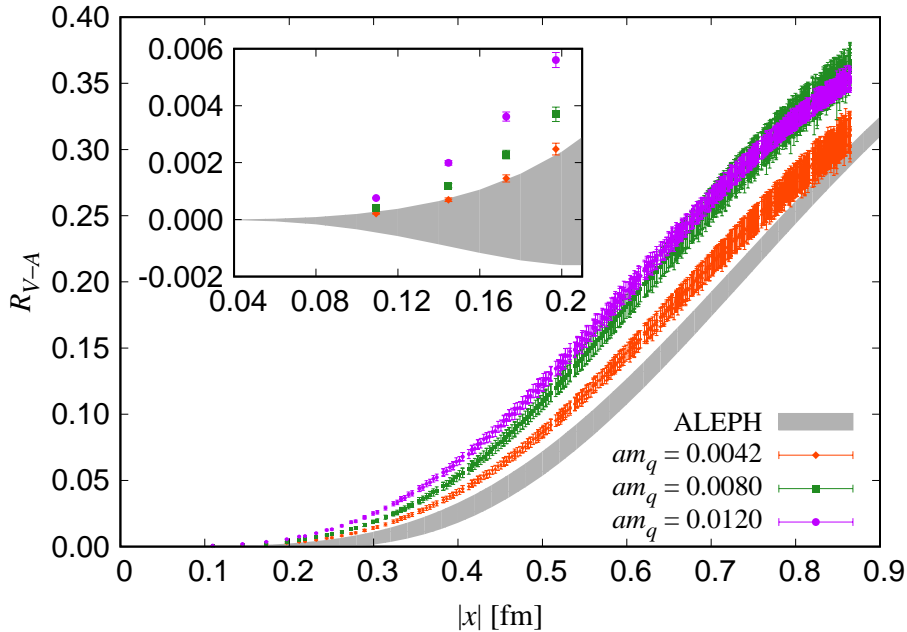}
    \caption{
      $R_{V-A}$ calculated on the ensembles with
      $\beta$ = 4.35, $am_s$ = 0.0180
      and three input light quark masses:
      $am_q$ = 0.0042 (diamonds), 0.080 (squares) and 0.0120 (circles).
      The results from experiment converted including the
      duality-violating term in (\ref{eq:spfunc_dv}) (band) are also
      shown. 
    }
    \label{fig:ALEPHvma}
  \end{center}
\end{figure}
\begin{figure}[tbp]
  \begin{center}
    \includegraphics[width=120mm]{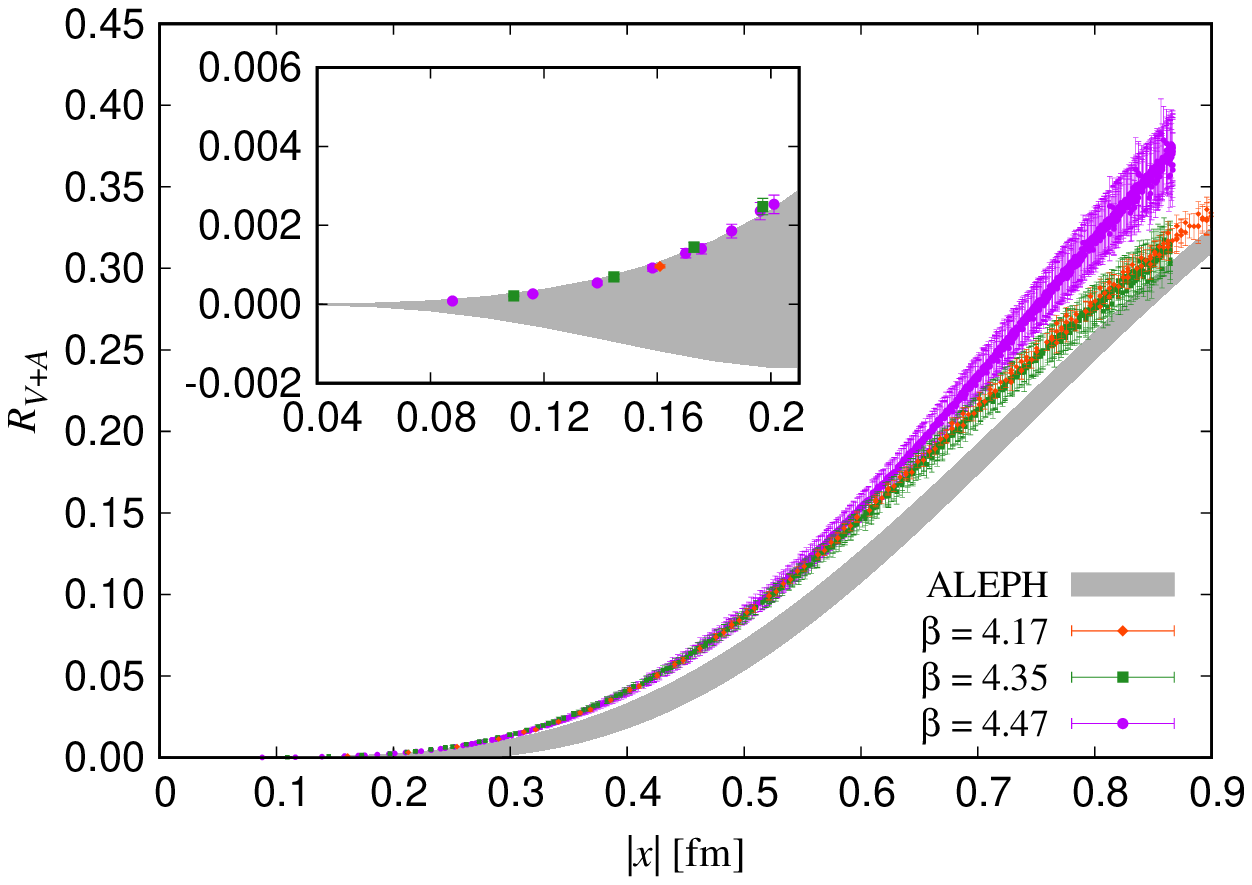}
    \caption{
      Same as Figure~\ref{fig:ALEPHvma} but calculated on 
      ensembles with different lattice cutoffs. Pion masses are
      $M_\pi\simeq$ 300~MeV. 
    }
    \label{fig:ALEPHvma_3beta}
  \end{center}
  \begin{center}
    \includegraphics[width=120mm]{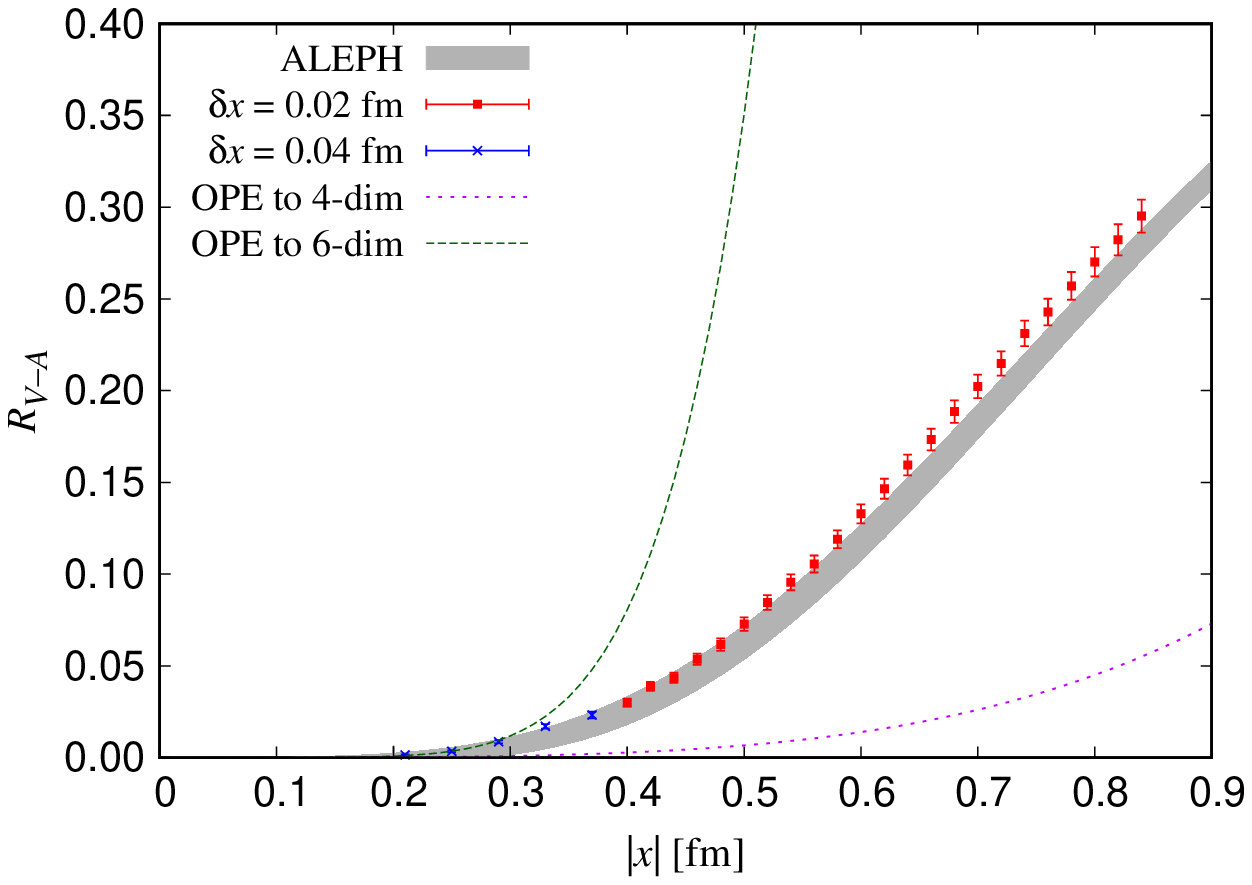}
    \caption{
      Lattice result for $R_{V-A}$ after the chiral and continuum
      extrapolations. 
      Data in each bin are extrapolated assuming (\ref{eq:extrap}).
      The bin size is larger in the short-distance region $|x|\lesssim$
      0.4~fm (blue crosses) than others (red squares) as there are fewer
      lattice points.
      The experimental result (band) and the predictions of the OPE
      including up to dimension-4 (dotted curve) and dimension-6 (dashed
      curve) operators are also plotted.
    }
    \label{fig:ALEPHvma_ext}
  \end{center}
\end{figure}

Next, we report the results for $R_{V-A}(x)$.
Figure~\ref{fig:ALEPHvma} shows the lattice data at three different
input masses at the same lattice spacing, $\beta=4.35$.
As expected, $R_{V-A}(x)$ vanishes in the short-distance limit because
of the good chiral symmetry of the M\"obius domain-wall fermion.
At short distances ($\lesssim$ 0.5~fm), the dependence on the input
mass is clearly seen and the results at smaller masses are closer to
the experimental result.

In Figure~\ref{fig:ALEPHvma_3beta}, which shows the results at
pion masses $M_\pi\simeq300$~MeV and at three different $\beta$'s,
there is no significant dependence on the lattice spacing
at short distances ($\lesssim$ 0.5~fm), unlike the case of the $V+A$
channel. 
One possible reason for this is that the discretization effect on
correlators at short distances is mostly perturbative and cancel
in the $V-A$ channel.

\begin{figure}[tbp]
  \begin{center}
    \includegraphics[width=120mm]{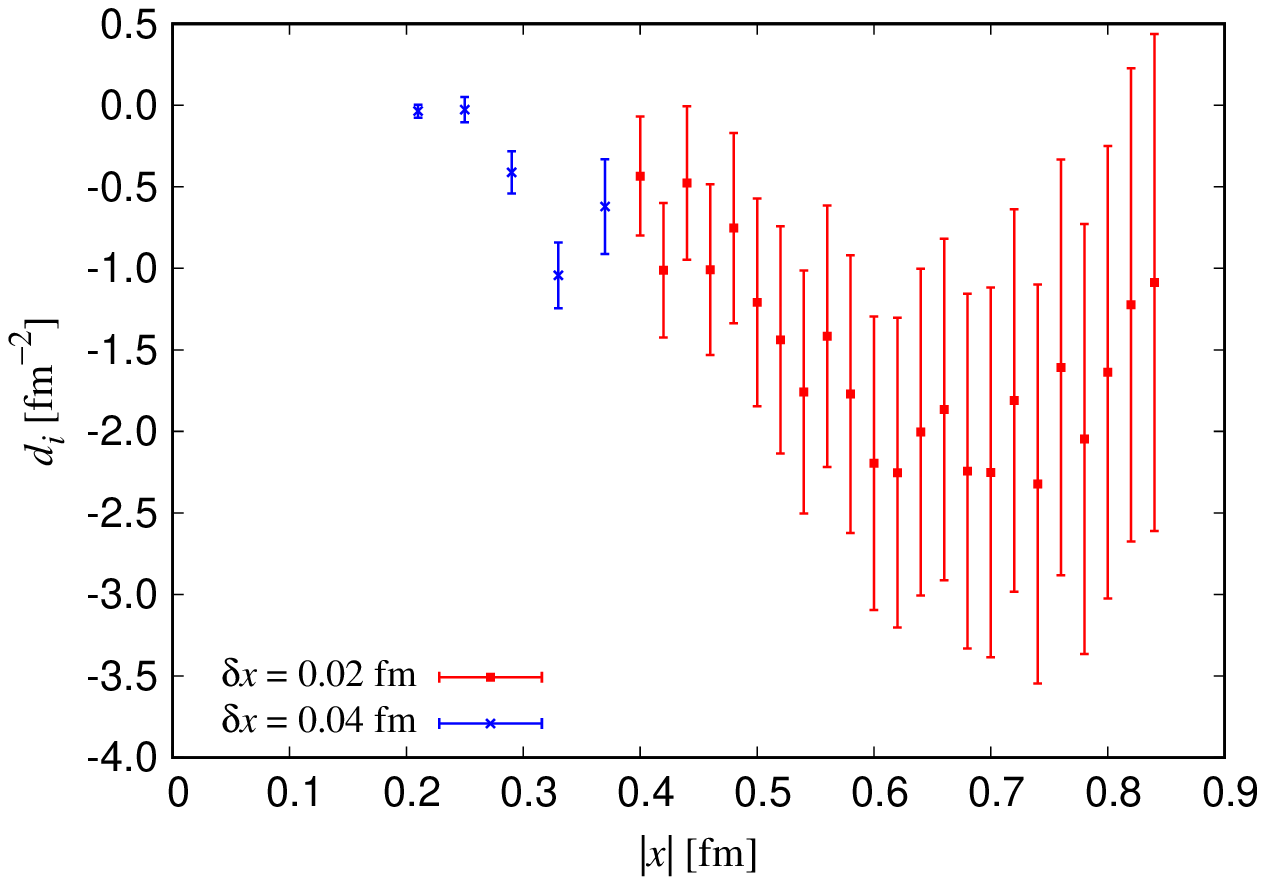}
  \end{center}
  \caption{
    Same as Figure~\ref{fig:ca_vpa} but for the $V-A$ channel.
  }
  \label{fig:ca_vma}
  \vspace{6mm}
  \begin{center}
    \includegraphics[width=120mm]{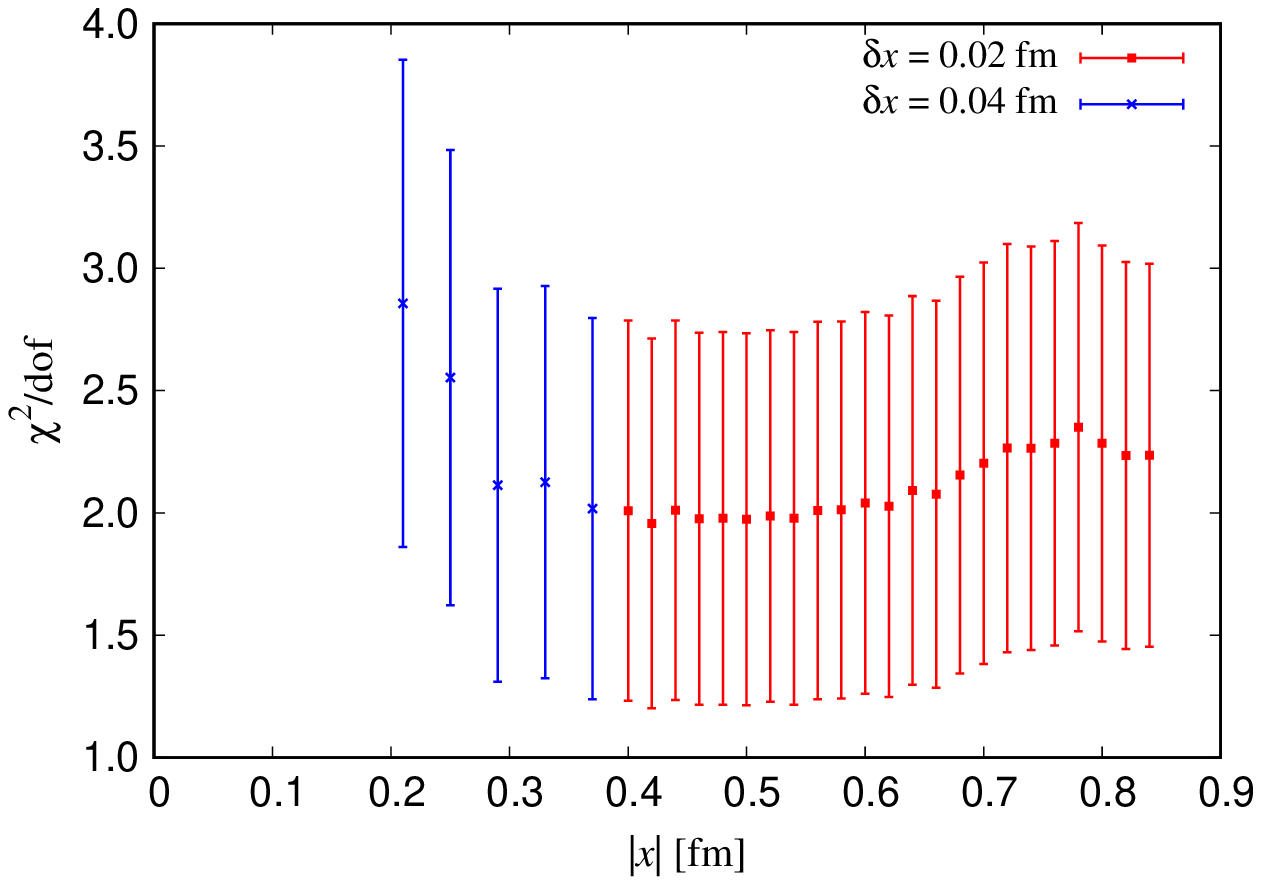}
  \end{center}
  \caption{
    Same as Figure~\ref{fig:chi2_vpa} but for the $V-A$ channel.
  }
  \label{fig:chi2_vma}
\end{figure}

\begin{figure}[tbp]
  \begin{center}
    \includegraphics[width=120mm]{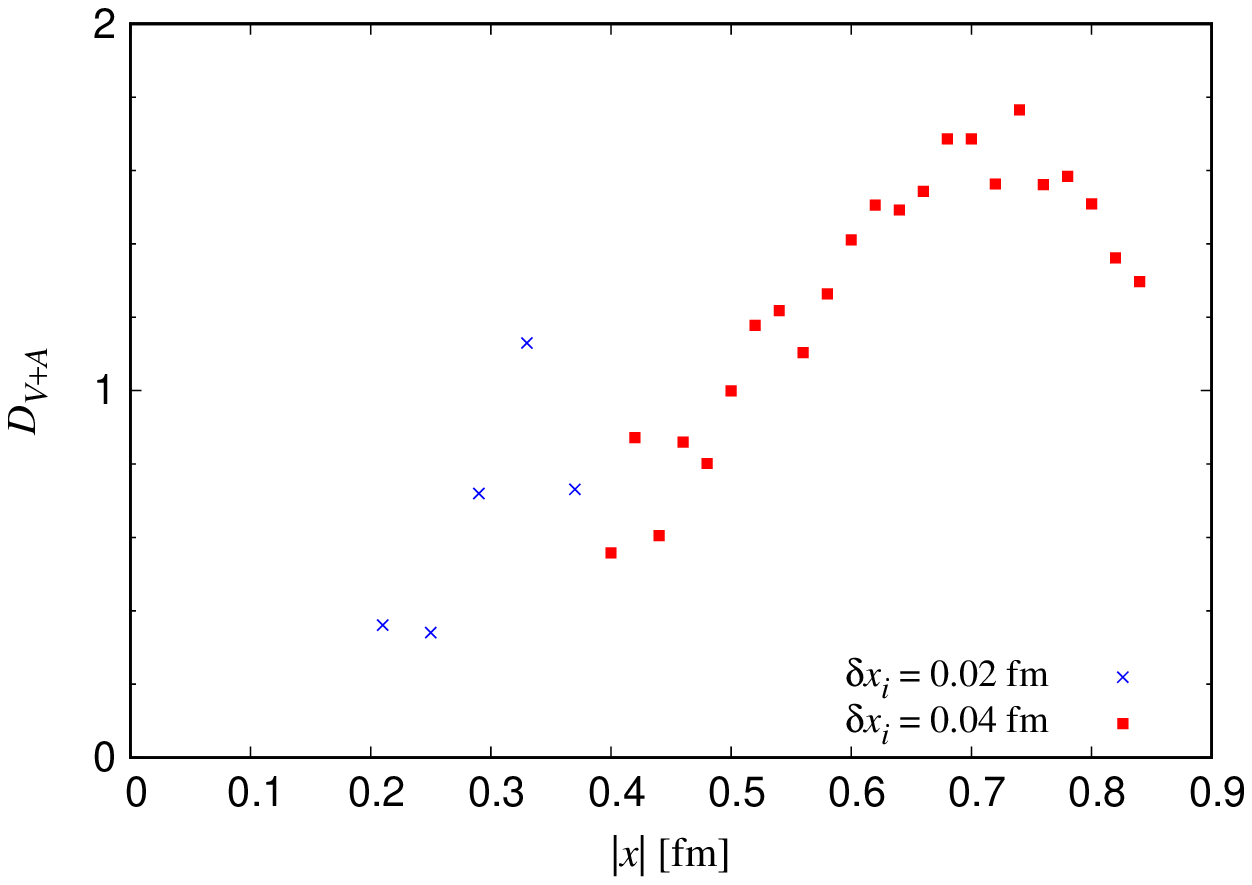}
    \caption{
	Significance of the difference $D_{V-A}(x)$ between the lattice calculation
	and experiment plotted as a function of $|x|$.
    }
    \label{fig:vma_sigma}
  \end{center}
\end{figure}

We extrapolate $R_{V-A}(x)$ to the physical point in the same manner
as for the $V+A$ channel.
The result is shown in Figure~\ref{fig:ALEPHvma_ext}.
The consistency between the lattice result and experiment can be seen
in the region $|x| >$ 0.2~fm.
The extrapolation formula works at $|x|\simeq0.2$~fm as
the fit result of $d_i$ and $\chi^2$/dof indicates (Figures~\ref{fig:ca_vma}
and \ref{fig:chi2_vma}).
Unlike the $V+A$ channel, $d_i$ does not vary rapidly
and $\chi^2$/dof remains $O(1)$ even at $|x|\simeq0.2$~fm.
At shorter distances there are few lattice points in a bin $B_i$ and the average
of $|x|$ over lattice points in the bin could significantly deviate from the center
$x_i$ of the bin depending on $a$.
This may lead to another source of $a$-dependence in
$\overline R_{V-A}(a,M_\pi,x_i)$, which may not be taken into account by our
fit procedure.
Although we do not extrapolate the lattice data in $|x|<0.2$~fm for this reason,
Figures~\ref{fig:ALEPHvma} and \ref{fig:ALEPHvma_3beta} indicate
that the lattice data agree with the phenomenological curve
even in the asymptotically small $|x|$ region.
Figure~\ref{fig:vma_sigma} shows the significance of the difference $D_{V-A}(x)$ between the
lattice calculation and experiment.
The largest difference, $\sim1.8\sigma$, is seen at $|x|=0.74$~fm, while
$1\sigma$ agreement is seen below $0.5$~fm, except for at $|x|=0.33$~fm,
where the fit function may not be appropriate as explained above.

Figure~\ref{fig:ALEPHvma_ext} also shows the predictions of the OPE
including up to dimension-4 (dotted curve) and dimension-6 (dashed
curve) operators.
The OPE $R_{V-A}^{\rm OPE}(x)$ of the $V-A$ channel is written as
\begin{equation}
  R_{V-A}^{\rm OPE}(x)
  = -\frac{\pi^2}{3}m_q\langle\bar qq\rangle x^4
  + \frac{\alpha_s\pi^3}{9}\langle\bar qq\rangle^2\ln (\mu_0x)^2x^6
  - \frac{f_\pi^2m_\pi^3\pi^2}{48}|x|^5K_1(m_\pi|x|)
  +O(m_q^2).
  \label{eq:OPE_V-A}
\end{equation}
Here, only the leading order of the strong coupling constant
$\alpha_s$ is shown for the first and second terms.
The first term is calculated by the Fourier transform of the OPE
in the momentum space given in \cite{Shifman:1978bx}.
The second term is estimated using the vacuum saturation approximation
with $\langle\bar qq\rangle^2$ \cite{Narison:2001ix}.
Before the normalization of (\ref{eq:Rvpma}),
this term is logarithmic in $x$ with an unknown parameter $\mu_0$
as a result of the Fourier transform
$Q^{-4}\rightarrow-{1\over16\pi^2}\ln(\mu_0x)^2$.
Since the first two terms on the right hand side of (\ref{eq:OPE_V-A})
correspond to the OPE including the longitudinal component of the
axial-vector correlators, we subtract the
contribution of the pion pole by the third term.
In the evaluation of $R_{V-A}^{\rm OPE}(x)$ shown in
Figure~\ref{fig:ALEPHvma_ext}, 
we set nominal values $f_\pi$ = 130~MeV, $m_\pi$ = 140~MeV,
and the scheme-dependent parameters at 2~GeV in the
$\rm\overline{MS}$ scheme, $\alpha_s$ = 0.3, $m_q$ = 3.4~MeV,
$\langle\bar qq\rangle = - \rm(270~MeV)^3$.
We also set $\mu_0=2$~GeV as a typical value.
The result of the truncation at dimension-4 already deviates from
the lattice result at 0.3~fm.
Including the dimension-6 operators, it still disagrees with the 
lattice result in the region $|x|> 0.3$~fm.

This analysis demonstrates the limitation of the operator expansion quantitatively.
The distance scale where the OPE can be safely used depends on the channel.
In the $V-A$ channel, it can only be safely used at $\lesssim$ 0.3~fm.

\subsection{Chiral condensate}
\label{subsec:efcc}

Here, we show our analysis to extract the chiral condensate through
the PCAC relation.
The basic recipe, which is valid in the continuum theory, is explained in
Section~\ref{subsec:PCAC_efcc}.

On the lattice, some modifications to (\ref{eq:AWI_cont}) are needed.
The violation of the current conservation induces substantial
discretization effects from the derivative term.
Such discretization effects can be largely
eliminated by subtracting the vector counterpart, which vanishes in
the continuum theory.

We analyze
\begin{equation}
  \Sigma_{m_q}^{V/A}(x) =
  -\frac{\pi^2}{2(m_q+m_{res})} x^2\sum_{\mu,\nu}x_\nu 
  \nabla_\mu \Pi_{V/A,\mu\nu}^\infty(x), 
  \label{eq:efcc_va}
\end{equation}
where the derivative $\nabla_\mu$ on the lattice is defined as
\begin{equation}
  \nabla_\mu f(x) = \frac{f(x+a\hat\mu)-f(x-a\hat\mu)}{2a},
\end{equation}
with $\hat\mu$ being the unit vector along the $\mu$-direction.
The residual mass $m_{res}$ is added to the quark mass to take
account of the violation of the Ginsparg-Wilson relation due to finite
$L_s$. 

In (\ref{eq:efcc_va}), we use the correlators
$\Pi_{V/A,\mu\nu}^\infty(x)$ after subtracting the finite volume effect.
Here, we assume that there is no significant finite volume effect for
the vector channel,
$\Pi_{V,\mu\nu}^\infty(x)=\Pi_{V,\mu\nu}(x)$,
which is justified because the single pion does not propagate in this
channel. 
The finite volume effect on the axial-vector channel is estimated as
the wrap-around effect of the pion as in the previous subsection.
The asymptotic form of 
$\sum_{\mu,\nu}x_\nu\nabla_\mu\Pi_{\rm A,\mu\nu}^\infty(x)$
at long distances is given by
\begin{equation}
  \sum_{\mu,\nu}x_\nu\nabla_\mu\Pi_{A,\mu\nu}^\infty(x)
  \rightarrow
  \frac{M_\pi^2z_0}{2\pi^2}\sum_\mu x_\mu\del_\mu\frac{K_1(M_\pi|x|)}{|x|},
  \label{eq:asym_xda}
\end{equation}
where $z_0$ and $M_\pi$ may be extracted from the zero-momentum
correlator, 
$\int\td^3x\ \Pi_{A,44}(\vec x,t) \rightarrow z_0\e^{-M_\pi t}$.
We can thus subtract the finite volume effect from the lattice data by
\begin{equation}
\sum_{\mu,\nu}x_\nu\nabla_\mu\Pi_{A,\mu\nu}^\infty(x)
= \sum_{\mu,\nu}x_\nu\nabla_\mu\Pi_{A,\mu\nu}(x)
-\frac{M_\pi^2z_0}{2\pi^2}\sum_{\mu,x_0}
x_\mu\del_\mu\frac{K_1(M_\pi|x-x_0|)}{|x-x_0|},
\end{equation}
where the sum over $x_0$ runs over
\begin{equation}
x_0 \in \{(\pm L,0,0,0),(0,\pm L,0,0), (0,0,\pm L,0),(0,0,0,\pm T), (\pm L,\pm L,0,0),\ldots\}.
\end{equation}

\begin{figure}[tbp]
  \begin{center}
    \includegraphics[width=120mm]{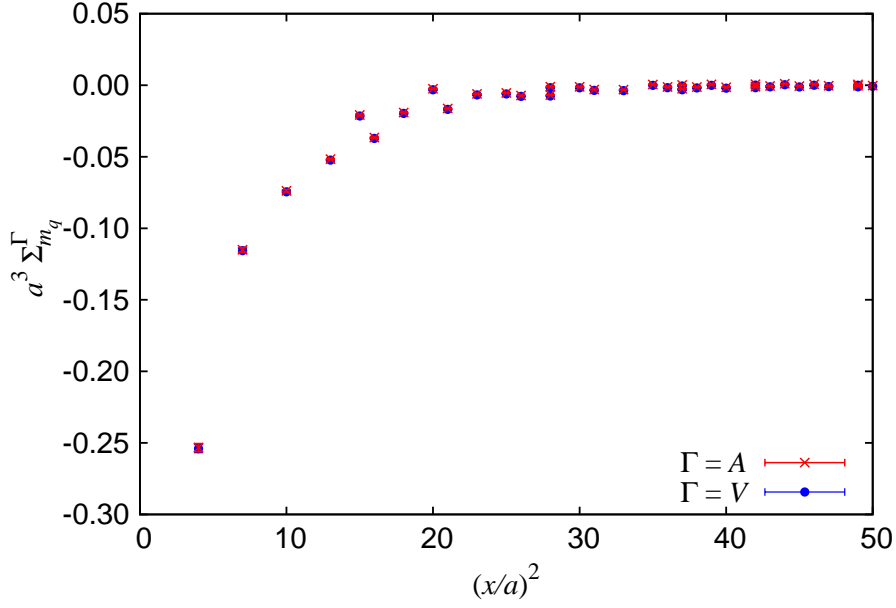}
    \caption{
      $\Sigma_{m_q}^{V/A}$ as functions of $(x/a)^2$ calculated on the
      ensemble with $\beta$ = 4.35 and 
      $(am_q,am_s) = (0.0042,0.0180)$.
    }
    \label{fig:efcc_nonsubt}
  \end{center}
\end{figure}

Figure~\ref{fig:efcc_nonsubt} shows the lattice result for 
$\Sigma_{m_q}^{V/A}(x)$
calculated on one of the ensembles.
In the continuum theory, the axial-vector channel $\Sigma_{m_q}^A(x)$
is equal to the chiral condensate up to the correction of $O(m_q)$, as
shown in (\ref{eq:AWI_cont}), and the vector channel is identically zero.
Figure~\ref{fig:efcc_nonsubt} indicates that 
the axial-vector channel calculated on the lattice almost coincides
with the vector channel and is non-zero, because of the non-conserving
(axial-)vector currents.
In other words, the axial-vector channel is largely contaminated by
discretization effects, by the same amount as in the vector
channel.
The precise chiral symmetry realized by M\"obius domain-wall fermion
is the source of this coincidence.

\begin{figure}[tbp]
  \begin{center}
    \includegraphics[width=120mm]{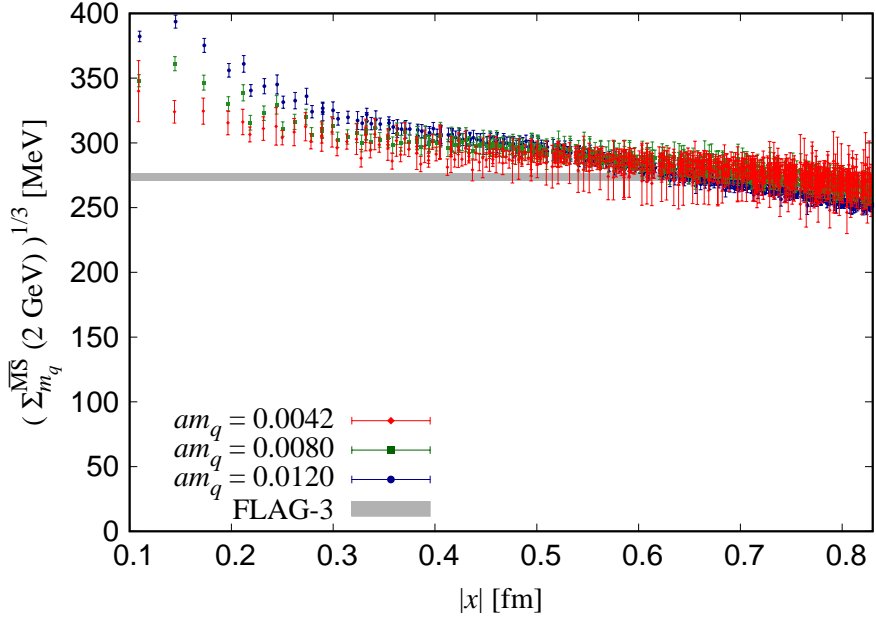}
    \caption{
      $(\Sigma_{m_q}^{\rm\overline{MS}}(2\rm\ GeV))^{1/3}$ calculated
      on the ensembles with
      $\beta$ = 4.35, $am_s$ = 0.0180
      and at three input light quark masses:
      $am_q$ = 0.0042 (diamonds), 0.080 (squares) and 0.0120 (circles).
      The gray band shows the FLAG average 
      $(\Sigma^{\rm\overline{MS}}(2{\rm~GeV}))^{1/3} = 274(3)$~MeV. 
    }
    \label{fig:efcc_ams0.0180}
  \end{center}
\end{figure}

\begin{figure}[tbp]
  \begin{center}
    \includegraphics[width=120mm]{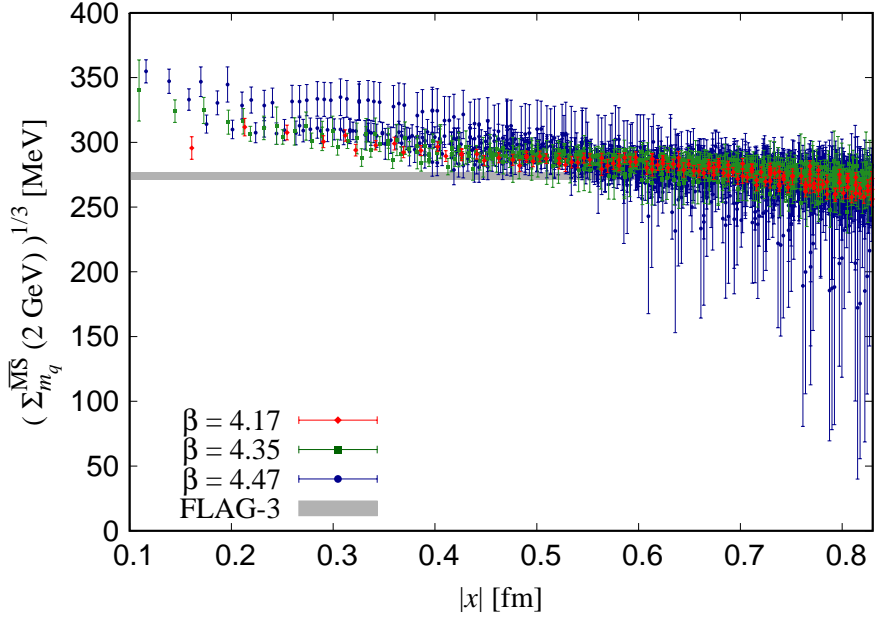}
    \caption{
      Same as Figure~\ref{fig:efcc_ams0.0180} but calculated on 
      ensembles with different lattice cutoffs and pion masses are
      $M_\pi\simeq$ 300~MeV. 
    }
    \label{fig:efcc_3beta}
  \end{center}
\end{figure}

Therefore, we may cancel the bulk of the discretization effects by
analyzing 
\begin{equation}
  \Sigma_{m_q}(x) = \Sigma_{m_q}^A(x) - \Sigma_{m_q}^V(x),
  \label{eq:efcc_a-v}
\end{equation}
which reduces to $\Sigma_{m_q}^A(x)$ in the continuum theory.
Figure~\ref{fig:efcc_ams0.0180} shows the cubic root of $\Sigma_{m_q}(x)$
calculated at three different masses and the same $\beta$.
The renormalization is done multiplicatively to the $\rm\overline{MS}$
scheme at 2~GeV,
\begin{equation}
  \Sigma_{m_q}^{\rm\overline{MS}}({\rm2~GeV};x)
  = {Z_V^{\rm\overline{MS}}(a)}^2 
  Z_S^{\rm\overline{MS}}(2~{\rm GeV};a) \Sigma_{m_q}(x)
\end{equation}
with the renormalization factors $Z_V^{\rm\overline{MS}}(a)$ and
$Z_S^{\rm\overline{MS}}(2~{\rm GeV};a)$ determined in the previous work
\cite{Tomii:2016xiv}.
The gray band represents the FLAG average \cite{Aoki:2016frl}
of the chiral condensate
$(\Sigma^{\rm\overline{MS}}({\rm 2\ GeV}))^{1/3}$ = 274$\pm3$~MeV
at $n_f=2+1$.

In Figure~\ref{fig:efcc_ams0.0180}, the results at smaller masses are
closer to the FLAG average.
This agrees with the theoretical expectation discussed in Section~\ref{subsec:PCAC_efcc}
that the chiral condensate 
$\Sigma^{\rm\overline{MS}}(\rm2~GeV)$ is obtained
in the chiral limit of $\Sigma_{m_q}^{\rm\overline{MS}}({\rm2~GeV};x)$.
Figure~\ref{fig:efcc_3beta} shows the results at three different
lattice spacings with pion masses $M_\pi\simeq300$~MeV.
Like for $R_{V-A}(x)$, there is no significant dependence on the
lattice spacing.

We extrapolate these results to the chiral and the continuum limits as
follows. 
The average 
$\overline\Sigma_{m_q}^{\rm\overline{MS}}({2\rm~GeV};a,M_\pi,x_i)$
of 
$\Sigma_{m_q}^{\rm\overline{MS}}({2\rm~GeV};x)$ 
over lattice points in each bin is defined similarly to 
$\overline R_{V\pm A}(a,M_\pi,x_i)$ in the previous subsections.
Unlike for $R_{V\pm A}(x)$,
the $x$-dependence of $\Sigma_{m_q}(x)$ is limited to
$O(M_\pi^2)$ or $O(a^2)$, and
the result of the extrapolation of $\Sigma_{m_q}(x)$
must be independent of $x$.
We therefore perform a simultaneous fit for all bins using the fit
function 
\begin{equation}
  \left(
    \overline\Sigma_{m_q}^{\rm\overline{MS}}({2\rm~GeV};a,M_\pi,x_i)
  \right)^{1/3}
  =
  \left( \Sigma^{\rm\overline{MS}}({2\rm~GeV}) \right)^{1/3} 
  + c_iM_\pi^2 + d_ia^2,
  \ \ \ \ 
  i = 1,2,\ldots, N,
  \label{eq:extrap_efcc}
\end{equation}
with $2N+1$ parameters
$c_1, c_2, \ldots, c_N, d_1, d_2, \ldots, d_N$
and
$\big(\Sigma^{\rm\overline{MS}}({2\rm~GeV})\big)^{1/3}$.

At short distances, the continuum extrapolation may be contaminated
because there are few data points in each bin.
At long distances, on the other hand, the extrapolation by the fit function
(\ref{eq:extrap_efcc}) may not be appropriate since the mass dependence
of $\Sigma_{m_q}(x)$ may be complicated, as described in (\ref{eq:asym_xda}).
We extrapolate lattice data at middle distances where the dependences
on the pion mass and lattice spacing would be well under control by the
fit function (\ref{eq:extrap_efcc}).

\begin{table}[t]
\caption{
Chiral condensate extracted from the global fit using (\ref{eq:extrap_efcc})
at various fit ranges and bin widths.
}
\label{tab:fit_efcc}
\begin{center}
\begin{tabular}{|cccc|c|}
\hline
$x_1-\delta x/2$ [fm] & $x_N+\delta x/2$ [fm] & $\delta x$ [fm] & $N$
& ${\Sigma^{\rm\overline{MS}}({2\rm~GeV})}^{1/3}$ [MeV] \\
\hline
0.23 & 0.83 & 0.02 & 30 & $284.3(4.0)$ \\[0mm]
0.23 & 0.83 & 0.04 & 15 & $285.2(4.0)$ \\[0mm]
0.23 & 0.83 & 0.06 & 10 & $284.5(4.0)$ \\[0mm]
0.23 & 0.83 & 0.10 &  6 & $285.7(4.0)$ \\[0mm]
\hline
0.23 & 0.43 & 0.04 & 5 & $293.7(5.4)$ \\[0mm]
0.31 & 0.51 & 0.04 & 5 & $290.7(5.0)$ \\[0mm]
0.39 & 0.59 & 0.04 & 5 & $288.9(4.6)$ \\[0mm]
0.47 & 0.67 & 0.04 & 5 & $285.5(4.4)$ \\[0mm]
0.55 & 0.75 & 0.04 & 5 & $280.8(4.0)$ \\[0mm]
0.63 & 0.83 & 0.04 & 5 & $276.2(3.8)$ \\[0mm]
\hline
\end{tabular}
\end{center}
\end{table}

Table~\ref{tab:fit_efcc} summarizes the results at several fit ranges
and widths of bins.
%
The dependence on $\delta x$ is sufficiently small compared to the statistical error.
On the other hand, the dependence on the fit range is larger than the statistical error.
Including this uncertainty in the estimate of the systematic error, we
determine the chiral condensate to be
\begin{equation}
  \big(\Sigma^{\rm\overline{MS}}(2\rm~GeV)\big)^{1/3}
  = 284.9 \pm 4.0_{\rm stat} \pm 8.8_{\rm sys}
  \mathrm{~MeV}.
\end{equation}
Here, the central value and the statistical error are estimated by an
average of the four results for the fit range 0.23--0.83~fm with
various widths $\delta x$,
while the systematic error is estimated as the maximum difference between
the central value and the results at various fit ranges.
This result agrees well with
the result obtained from the Dirac spectrum 
on the same set of lattice ensembles, $270.0(4.9)$~MeV
\cite{Cossu:2016eqs}. 

\section{Conclusion}
\label{sec:summary}

We have discussed the vector and axial-vector current correlators in
the distances between the perturbative and non-perturbative regimes.
In this intermediate region, neither the perturbative approaches nor
low-energy effective theories are fully applicable.
Lattice calculation can be used to analyze such theoretically
difficult physical regions, provided that the systematic errors are
properly estimated.
The $\tau$ decay experiment by ALEPH played a crucial role as it
provides the data for both the vector and axial-vector channels.
The $V+A$ channel is mainly useful to test the perturbation theory,
while the $V-A$ is sensitive to the non-perturbative aspects of QCD.
The lattice calculation of the $V+A$ channel agrees with experiment
in the length scale of $|x|\ge0.4$~fm at the level of $1.3\sigma$ or better.
The $V-A$ channel agrees at a level of $1.8\sigma$ at $|x|=0.74$~fm
or even better agreement at other distances.
The chiral condensate, the order parameter of chiral symmetry
breaking, is also precisely extracted from this analysis.

The consistency between the lattice calculation and the experimental
data seen in this work adds further support for the validity of QCD in
the distance region where excited states contribute significantly.
The lattice calculation of the current correlators can also identify
the region where the correlators are well explained by the OPE, which is
the main theoretical tool in phenomenological analyses.

The method applied in this work to extract the chiral condensate 
needs a differential of the axial-vector correlator on the lattice.
The substantial discretization error from the differential is
dramatically reduced by subtracting the counterpart of the vector
channel. 
The fact that the obtained chiral condensate agrees with the result based
on the Banks-Casher relation gives additional confidence in our
description of the symmetry broken QCD vacuum.

Lattice calculations of current correlators at finer lattices, which
will become available in the near future, would be very interesting as
they may give further information in the perturbative regime.
They will provide us with stringent tests of the perturbative expansion
in QCD, which is now available to $O(\alpha_s^4)$ for the vacuum
polarization function.
At the same time, they would be a sensitive probe to determine the strong
coupling constant.

While this work focused on the vector and axial-vector channels,
the scalar and pseudoscalar correlators may reflect a different aspect
of QCD.
The perturbative region for these correlators is much shorter than
that of the vector and axial-vector channels because they are directly
affected by instanton interactions \cite{Novikov:1981xi}.
Lattice calculations of the scalar and pseudoscalar correlators
may give new insights into understanding such effects.

\begin{acknowledgments}
Numerical simulations are performed on the Hitachi SR16000 and IBM 
Blue Gene/Q at KEK under a support of its Large Scale
Simulation Program (No.~15/16-09, 16/17-14).
We thank P. Boyle for providing the highly optimized code for Blue
Gene/Q.
We also thank B.~Colquhoun for careful reading of the manuscript.
This work is supported in part by the US DOE grant (\#DE-SC0011941),
the Grant-in-Aid of the Japanese Ministry of Education
(No.~25800147, 26247043, 26400259)
and the Post-K supercomputer project through JICFuS.
\end{acknowledgments}

\bibliography{SDC}
\end{document}